\documentclass[useAMS,usenatbib]{mn2e}  
\usepackage{graphicx,epsfig,psfig}  
\usepackage{booktabs} 
\usepackage{kellymacros} 
\usepackage[usenames]{color}

\long\def\Ignore#1{\relax}

\title[KIC\,4544587: an Eccentric Binary with $\delta$ Sct Pulsations]{KIC 
4544587: an Eccentric, Short Period Binary System with $\delta$ Sct Pulsations 
and Tidally Excited Modes} 
 
\author[Hambleton et~al.]{K. M. Hambleton$^{1,2,3}$\thanks{Email:  
kmhambleton$@$uclan.ac.uk}, D. W. Kurtz$^{1}$, A. Pr{\v s}a$^{3}$, J. A. 
Guzik$^{4}$, K. Pavlovski$^{5}$, \newauthor S. Bloemen$^{2}$, J. 
Southworth$^{6}$, K. Conroy$^{7}$, S. P. Littlefair$^{8}$ and J. Fuller$^{9}$\\ 
$^{1}$Jeremiah Horrocks Institute, University of Central Lancashire, Preston, 
PR1~2HE\\ 
$^{2}$Instituut voor Sterrenkunde, KU Leuven, Celestijnenlaan 200D, B-3001 
Leuven, Belgium\\ 
$^{3}$Department of Astronomy and Astrophysics, Villanova University, 800 East 
Lancaster Avenue, Villanova, PA 19085, USA\\  
$^{4}$Los Alamos National Laboratory, XTD-2 MS T-086, Los Alamos, NM 87545-2345, 
USA\\ 
$^{5}$Department of Physics, Faculty of Science, University of Zagreb, Croatia\\ 
$^{6}$Astrophysics Group, Keele University, Staffordshire,ST5 5BG\\ 
$^{7}$Department of Physics and Astronomy, Vanderbilt University, Nashville, TN 
37235\\ 
$^{8}$Department of Physics and Astronomy, University of Sheffield, Sheffield S3 
7RH\\ 
$^{9}$Center for Space Research, Department of Astronomy, Cornell University, Ithaca, NY 14853, USA\\ 
}

\begin{document} 
\date{Accepted}  
\pagerange{\pageref{firstpage}--\pageref{lastpage}} \pubyear{2011}  
 
\maketitle  
\label{firstpage}

\begin{abstract}

\noindent 
We present \Kep photometry and ground based spectroscopy of KIC\,4544587, a 
short-period eccentric eclipsing binary system with self-excited pressure and 
gravity modes, tidally excited modes, tidally influenced p\,modes, and rapid 
apsidal motion of 182\,y per cycle. The primary and secondary components of KIC\,4544587 reside within the \DS and \GD instability region of the Hurtzsprung-Russell diagram, respectively. By applying the binary modelling software 
{\sc phoebe} to prewhitened \Kep photometric data and radial velocity data 
obtained using the William Herschel Telescope and 4-m Mayall telescope at KPNO, 
the fundamental parameters of this important system have been determined, 
including the stellar masses, 1.98\,$\pm$0.07\,$\Msun$ and 
1.60\,$\pm$\,0.06\,$\Msun$, and radii, 1.76\,$\pm$\,0.03\,\Rsun\, and 
1.42\,$\pm$\,0.02\,\Rsun, for the primary and secondary components, 
respectively. Frequency analysis of the residual data revealed 31 modes, 14 in 
the gravity mode region and 17 in the pressure mode region. Of the 14 gravity 
modes 8 are orbital harmonics: a signature of tidal resonance. While the 
measured amplitude of these modes may be partially attributed to residual signal from binary model subtraction, we demonstrate through consideration of the 
folded light curve that these frequencies do infact correspond to tidally excited pulsations. Furthermore, 
we present an echelle diagram of the pressure mode frequency region (modulo the 
orbital frequency) and demonstrate that the tides are also influencing the 
p\,modes. A first look at asteroseismology hints that the secondary component is responsible for the p modes, which is contrary to our expectation that the hotter star should pulsate in higher radial overtone, higher frequency p modes.

\end{abstract} 
 
 
\section{Introduction}  
\label{sec:intro} 
 
The \DS stars form an integral part of the instability strip, spanning a 2-mag range of evolutionary stages, from pre-main sequence to the terminal-age main sequence \citep{Rodriguez2001}. Their luminosities range from $0.6 \le 
\log(L/ \Lsun) \le 2.0$ and their effective temperatures range from $6300 \le T_{\rm  
eff} \le 9000$\,K \citep{Buzasi2005}. They oscillate in radial and non-radial 
pressure modes (p\,modes) and low order gravity modes (g\,modes) with observed 
periods ranging from approximately 18\,min to 8\,h 
\citep{Amado2004,Pamyatnykh2000,Grigahcene2010b}. 
 
\begin{figure*} 
\hfill{} 
\includegraphics[width=\hsize, height=10cm]{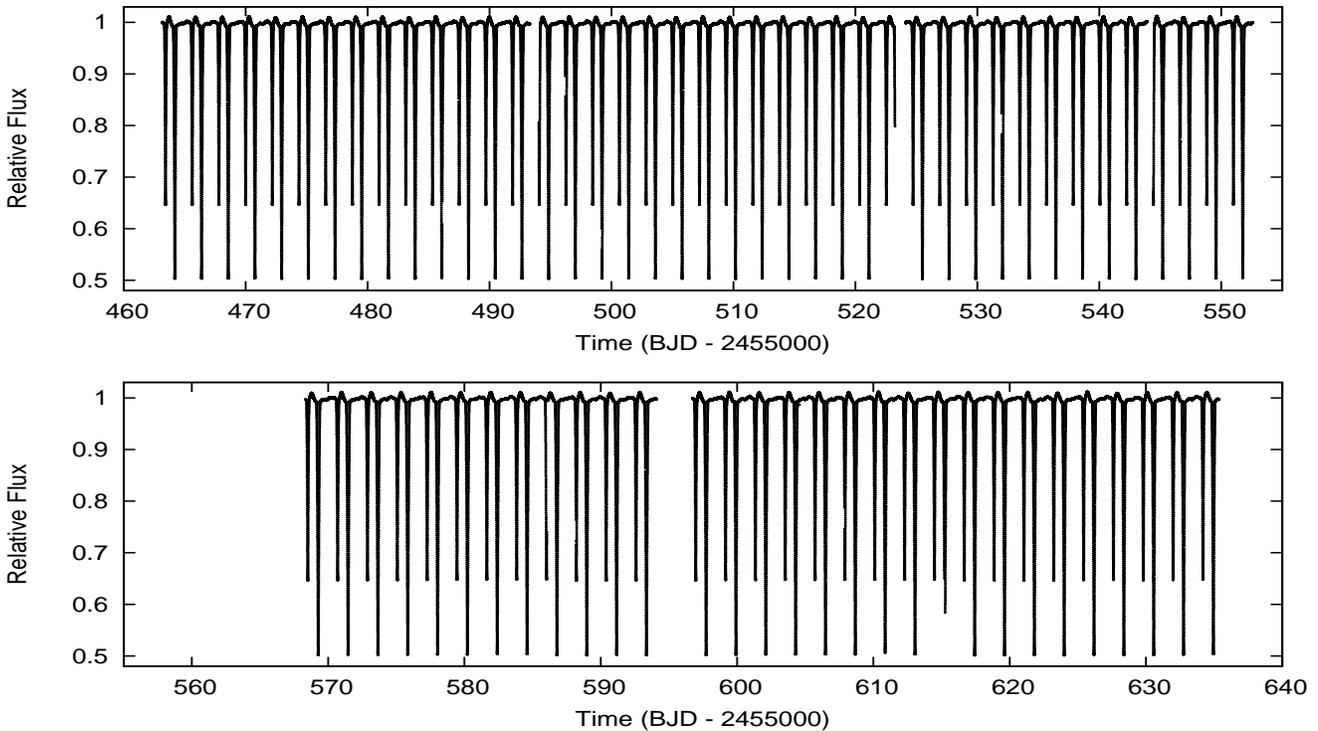}\\ 
\small\caption{The observed {\it Kepler} short cadence light curve of 
KIC\,4544587 for Quarters\,7 (upper panel) and 8 (lower panel). Data are missing 
from the beginning of Quarter\,8 due to a safe-mode event. The time is in BJD.} 
\label{LC} 
\hfill{} 
\end{figure*}

The $\kappa$ mechanism is the primary driving mechanism of \DS pulsations, 
although \citet{Antoci2011} suggested that one \DS star may pulsate with 
stochastically excited modes similar to those seen in the Sun and solar-like 
pulsators. Delta Scuti stars have a mass range between 1.5 and 2.5\,$\Msun$ 
\citep{Lefevre2009}. At approximately 2\,$\Msun$ there is a transitional phase 
where the size of the convective outer envelope  becomes negligible for higher 
mass stars and their outer envelopes become dominated by radiative energy 
transport; at approximately 1.5\,$\Msun$, stars of higher mass develop a 
convective core \citep{Aerts2010}. As this critical transition in the convective 
envelope occurs within the range of masses encompassed by \DS stars, the 
asteroseismic investigation of \DS stars may eventually unveil fundamental 
information pertaining to the physical processes that govern this transition. 
 
Gamma Dor stars are main sequence stars in the temperature range  $6800 \le 
T_{\rm  eff} \le 7600$\,K that pulsate in high order gravity modes driven by 
convective blocking \citep{Guzik2000} with pulsation periods typically of the 
order of 1\,d \citep{Grigahcene2010a}. As \GD stars are a relatively new class 
of stars \citep{Balona1994}, until recently their recorded numbers were low and 
consequently \DS stars were believed to dominate the classical pulsators on the 
main sequence \citep{Breger2000}. However, with the implementation of advanced 
instruments such as {\it Kepler} \citep{Borucki2010, Gilliland2010, 
Batalha2010}, {\it MOST} \citep{Walker2003} and {\it CoRoT} \citep{Baglin2006}, 
many stars demonstrating both \DS and $\gamma$ Dor characteristics have been 
observed; thus new classification criteria, containing $\gamma$\,Dor--\DS and 
$\delta$ Sct--$\gamma$ Dor hybrid stars, have been introduced 
\citep{Grigahcene2010b}. Following this revision, through the characterisation 
of 750 A-F type main-sequence stars, the percentage of \DS stars on the main 
sequence is now estimated to be 27\, per cent, \GD stars accounting for 13 per cent, hybrids 
accounting for 23 per cent and the remaining stars being classified as other types of 
variables, \eg spotted stars showing rotational variations \citep{Uytterhoeven2011}. 
 
In a study of 119 A0--A9 stars, 35\,$\pm$\,5 per cent were found to be in 
multiple systems \citep{Abt2009}. However, only 22 per cent of the \DS stars 
catalogued are known to be multiple stars \citep{Rodriguez2001}. In binary 
systems the rotational velocity of each stellar component tends towards a 
velocity that is synchronous with the orbital period as the orbit evolves. As 
synchronous velocity depends linearly on radius and scales with orbital period 
(to the power of 2/3), it generally implies an equatorial velocity less than 
of 120\,km\,s$^{-1}$ \citep{Abt2009}, while most $\delta$ Scuti stars are 
found to have velocities greater than 120\,km\,s$^{-1}$. Below this value the 
turbulence in the outer stellar envelope only enables a negligible amount of 
meridional mixing to occur. This allows for diffusion to take place, which prevents 
pulsation through the settling of Helium out of the He\,\textsc{ii} 
\citep{Breger1970}. Thus it has previously been assumed that multiplicity 
indirectly inhibits pulsation. However, there are many cases, including 
HD\,174884 \citep{Maceroni2009}, HD\,177863 \citep{Willems2002}, and the newly 
identified class of eccentric ellipsoidal variables known as \HB stars 
\citep{Thompson2012}  -- including the iconic KOI--54 \citep{Welsh2011, 
Fuller2012, Burkart2012} -- that demonstrate how, in some circumstances, 
multiplicity can not only alter, but increase pulsation amplitudes through the 
tidal excitation of eigenmodes. It is worthy of note, however, that the 
existence of tidally driven modes in binary star systems does not invalidate the 
theory that binarity also indirectly suppresses self-excited modes.   
 
Through the use of binary modelling techniques, direct measurements of stellar 
masses, radii and distances are possible. Asteroseismic modelling of the 
identified modes can provide information pertaining to the internal stellar 
structure and rotation of the pulsating component, making multiple systems with \DS components 
extremely valuable. Currently, the thorough asteroseismic analysis of \DS stars 
is rarely achieved due to our current inability to model a large number of 
oscillatory modes excited via the $\kappa$-mechanism. However, with the advent 
of cutting-edge observations from instruments such as \Kep and {\it CoRoT} and 
new methodologies such as those used by \citet{Garcia2009} on HD\,174936, it is 
expected that an increasing number of these intriguing objects will be solved in the foreseeable future.

\begin{table} 
\caption{ 
\label{tab:ID} 
Other identifiers and basic data for KIC\,4544587. The Kp passband specified 
is derived from the \Kep broadband filter.}  
\begin{center} 
\begin{tabular}{l l} 
\hline 
\multicolumn {2}{l}{Identifiers}\\\hline 
TYC                    & \ \ 3124-1348-1\\ 
GSC                    & \ \ 03124-01348\\ 
2MASS                  & \ \ J19033272+3941003\\\hline 
\multicolumn {2}{l}{Position and Brightness}\\\hline 
RA \ (J2000)           & \ \ 19:03:32.7274\\ 
Dec (J2000)            & +39:41:00.314\\ 
{\it V}                & \ \ 10.8\\ 
{\it B}                & \ \ 10.9\\ 
Kp			 		   & \ \ 10.8\\ 
\hline 
\end{tabular} 
\end{center} 
\end{table} 
 
The \Kep satellite, with its highly precise photometry, is generating 
observations unparallelled in precision and subsequently giving greater insight 
into the study of stellar structure through the use of asteroseismology. The 
primary objective of the \Kep mission is the identification and classification 
of planets through the transit method. However, the instrumentation required for 
such observations is highly applicable to the field of asteroseismology 
\citep{Gilliland2010}. Attributes, such as a stable platform, that enable 
extended observations, and a precision as good as a few parts per million make 
the \Kep observations quintessential for the advancement of asteroseismology. A 
dynamic range of over 10 mag, in addition to 
a 105\,deg$^2$ field of view, give \Kep an unprecedented advantage for obtaining 
high quality asteroseismic data. Moreover, the ability to generate short 
cadence data of $\sim$1\,min time resolution allows for detailed photometric 
analyses of pulsating stars across the H-R diagram. 
 
KIC\,4544587 (where KIC is an acronym for `\Kep Input Catalogue') is an 
eccentric ($e = 0.28$), short-period ($P = 2.1891$\,d) binary system that 
contains at least one pulsating component (cf. Table\,\ref{tab:ID} for a list of 
observable information and identifiers). It was initially identified as a binary 
by \citet{Prsa2011} as part of the first release of the Kepler Eclipsing Binary 
Catalog (http://keplerebs.villanova.edu). The temperature of the primary 
component is equivalent to a late A-type star that is within the \DS instability 
strip and the secondary component's temperature is indicative of an early F 
star, which is likely to be a $\gamma$ Dor variable. 

Primarily this object was selected as a likely candidate for tidally induced 
pulsations due to the close proximity of the components at periastron, 
$\sim$\,4\,\Rsun\ surface-to-surface. KIC\,4544587 also has interesting orbital 
characteristics including a brightening at periastron in the \Kep photometric 
light curve due to the combination of tidal distortion and sub-stellar heating. 
Such a feature is indicative of an eccentric binary with its components in close 
proximity \citep{Maceroni2009, Thompson2012}. 
 
In this paper information obtained from modelling the binary features of the 
photometric and radial velocity curves of KIC\,4544587, and the results of the 
pulsational frequency analysis are presented. In \S\ref{sec:Obs} the 
observations are discussed, including adjustments to the original data set. 
Sections\,\ref{sec:SpecDis} and \ref{sec:AtmoPar} describe the spectral 
disentangling and the atmospheric parameters determined from the disentangled 
spectra. In \S\ref{sec:Period} the determination of the orbital period is 
detailed. In \S\ref{sec:Mod} the binary light curve modelling method is 
discussed, which focuses on the use of the binary modelling software, {\sc 
phoebe} \citep{Prsa2005}. Section~\ref{sec:Pulse} contains the frequency 
analysis and includes discussion of the evidence for resonance effects. A 
summary of this paper, with concluding remarks, is given in 
\S\ref{sec:Conc}. 
 
\section{Observations} 
\label{sec:Obs} 
 
\subsection{{\it Kepler\/} photometry} 
 
\begin{table} 
\caption{ 
\label{tab:quarters}\small The number of data points and duty cycle acquired for 
each individual \Kep Quarter. The long cadence data (LC) correspond to a 
sampling rate of 29.4244\,min and short cadence data (SC) to a sampling rate of 
58.8488\,s.}  
\begin{center} 
\begin{tabular}{||l|l|l|l||} 
\hline 
Quarter & Cadence & Number of data points & Duty cycle\\\hline 
0       & LC      & 476                      & 99.5\%\\ 
1       & LC      & 1\,639                   & 98.1\%\\ 
3.2     & SC      & 44\,000                  & 98.5\%\\ 
6       & LC      & 4\,397                   & 97.2\%\\ 
7       & SC      & 128\,830                 & 98.1\%\\ 
8       & SC      & 98\,190                  & 94.3\%\\ 
\hline 
\end{tabular} 
\end{center} 
\end{table} 
 
\begin{figure*} 
\hfill{} 
\includegraphics[width=\hsize, height=7cm]{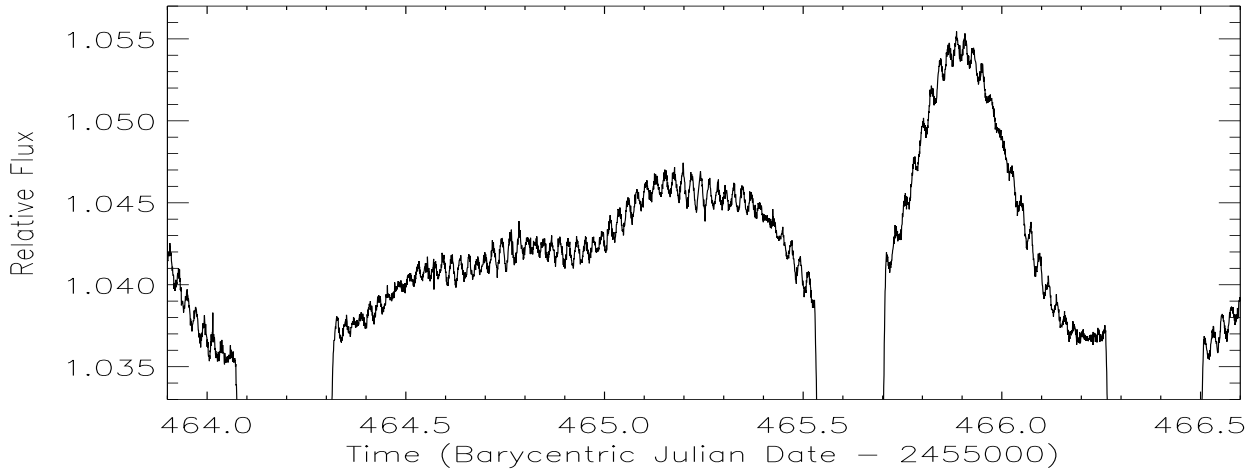}  
\small\caption{An amplified image of the out-of-eclipse phase of the {\it 
Kepler} Quarter\,7 SC light curve. Here the both the p-mode (periods in the 
range 30\,min -- 1\,hr) and the g-mode pulsations ($\sim$1\,d) are clearly 
visible. The pronounced periastron brightening can also be seen at approximately 
BJD\,2455466.} 
\label{ZoomLC} 
\hfill{} 
\end{figure*} 
 
The \Kep photometric observations of KIC\,4544587 consist of both long cadence 
(hereafter LC) data, during Quarters 0--11, and short cadence (hereafter SC) data 
during Quarters 3.2, 7, 8, 9 and 10. For our purposes we used a subset of these 
data up to and including Quarter\,8 (see Table\,\ref{tab:quarters} and 
Fig.\,\ref{LC}), which were available at the time of analysis. A Quarter is 
defined as a quarter of a complete, 372.5-d, \Kep orbit around the Sun 
\citep{Kjeldsen2010}. LC data correspond to a sampling rate of 29.4244\,min and 
SC data to a sampling rate of 58.8488\,s. For both formats 6.02-s exposures are 
co-added on board; this occurs 270 times to form an LC and 9 times to form an SC 
data point \citep{Caldwell2010}, with any remaining time attributed to readout 
time. The data are time-stamped with truncated Barycentric Julian date 
\citep{Gilliland2010}, which is Barycentric Julian date minus 2\,400\,000. The 
total \Kep photometric observations that have been analysed span from 2009 May--2011 March and comprise 277\,514 data points.  
 
The photometric observations were made using the \Kep broadband filter, which is 
similar to Cousins $Rc$. It is advised in the data release notes 
that accompany the \Kep data that the corrections made in the pipeline can have 
adverse affects on the binary signal in the data. For this reason the Simple 
Aperture photometry light curves were used instead of those created by the PPA 
(Photometer Performance Assessment) portion of the pipeline \citep{Li2010}.  
 
From the total data set, 9471 points were removed as outliers, of which 661 data 
points were removed from Quarter\,3.2, 2\,448 from Quarter\,7 and 5\,597 from 
Quarter\,8. These outliers were selected by eye as the intrinsic variations in 
the data significantly reduce the effectiveness of automated sigma clipping. 
Cosmic rays and other noise sources are the dominant causes of outliers and 
small gaps in the data are also present due to safe mode events, spacecraft 
rolls and brightening events known as Argabrightening, named after the 
discoverer, V. Argabright \citep{VanCleve2009}. These gaps, however, are 
minimal, which can be seen by the high duty cycle that was obtained for each 
Quarter independently, with the exception of the safe mode event at the 
beginning of Quarter\,8.  
 
The SC data have the advantage of increased time resolution, which enables the 
identification of the p-mode pulsations present in this object (see 
Fig.\,\ref{ZoomLC}). Consequently, Quarters 7 and 8 were used for the binary 
modelling and mode identification of KIC\,4544587 (with the exception of 
modelling the rate of apsidal advance where all LC data were used). A customised 
target mask was constructed for Quarters\,7 and 8 so that the average flux level 
was consistent over the two quarters. This is important for asteroseismic 
analysis as quarter-to-quarter flux variations  can cause instrumental amplitude 
modulation in the data. pyKE software (provided by the \Kep Guest Observer 
office: http://keplergo.arc.nasa.gov/) was used to define the mask, generate the 
new data files, and convert the data from {\sc fits} to {\sc ascii} format. We 
detrended and normalised each month of data individually by fitting a first or 
second order Legendre polynomial to segments of data separated by gaps (\ie 
caused by spacecraft rolls and safe mode events). As the eclipses affect the 
detrending process, we elected to fit the polynomials to the out-of-eclipse 
envelope only. The out-of-eclipse envelope was identified by sigma clipping the 
data. Using this method we (temporarily) removed all data points 5$\sigma$ above 
and 0.05$\sigma$ below the light curve. The long-term 
trends in the out-of-eclipse envelope were then fitted and the trends removed 
from the original data. Finally, we applied a Fourier transform to the 
polynomials and found that all significant peaks were $\nu < 0.03$\,d$^{-1}$, 
showing that we did not remove any information intrinsic to the system through 
this method.  
 
As each \Kep pixel is 4\,$\times$\,4\,arc\,seconds, it is expected that some 
contamination may occur within the photometric mask. The contamination value for 
KIC 4544587, specified by the Kepler Asteroseismic Science Operations Centre 
(KASOC), is estimated to be 0.019, where 0 implies no contamination and 1 
implies complete contamination of the CCD pixels. This contamination value 
suggests that KIC\,4544587 suffers minimally from third light, if at all. We 
applied the pyKE software to the target pixel files to assess the flux incident 
on each individual pixel. Light curves for each pixel were generated and the 
flux distribution over the newly defined masks were examined. From this we 
determined that the contamination level for KIC\,4544587 is negligible. 
 
\subsection{Ground-based spectroscopy} 
\label{sec:spectroscopy} 
 
\begin{table} 
\caption{ 
\label{tab:RVs} 
\small Radial velocity data of the primary (RV1) and secondary (RV2) components 
and their respective uncertainties (standard deviation) for 38 spectra obtained 
with the WHT and 5 spectra from the 4-m Mayall telescope. The Intermediate 
dispersion Spectrograph and Imaging System (ISIS) was used in conjunction with 
the WHT to obtain simulations red band ($6100 - 6730$\,\AA) and the blue band 
($4200 - 4550$\,\AA) spectra. The average radial velocity for each given time is 
specified. The echelle spectrograph was used on the 4-m Mayall telescope.} 
\begin{center} 
\begin{tabular}{l r r} 
\hline 
\multicolumn{1}{c}{Time} & \multicolumn{1}{c}{RV1}  & \multicolumn{1}{c}{RV2} \\ 
\multicolumn{1}{c}{(BJD)} & \multicolumn{1}{c}{ km\,s$^{-1}$} 
&\multicolumn{1}{c}{ km\,s$^{-1}$} \\ 
\hline 
\multicolumn {3}{c}{WHT}\\\hline 
2455730.62152	&	71.9	$\pm$	5.2	&	-133.8	        $\pm$	6.5
	\\ 
2455730.65750	&	91.2	$\pm$	4.7	&	-154.5	        $\pm$	5.8
	\\ 
2455730.69932	&	102.2	$\pm$	4.1	&	-177.3	        $\pm$	5.3
	\\ 
2455731.55625	&	-52.2   $\pm$	5.3	&	30.9		$\pm$	5.7	\\ 
2455731.60121	&	-62.3	$\pm$	5.3	&	37.8		$\pm$	5.7	\\ 
2455731.64668	&	-69.4	$\pm$	5.0	&	43.8		$\pm$	5.5	\\ 
2455731.70116	&	-77.3	$\pm$	4.8	&	50.4		$\pm$	5.9	\\ 
2455732.43958	&	-79.5	$\pm$	4.9	&	54.9		$\pm$	5.8	\\ 
2455732.48500	&	-67.9	$\pm$	4.7	&	46.0		$\pm$	5.3	\\ 
2455732.52538	&	-54.9	$\pm$	4.7	&	36.8		$\pm$	5.9	\\ 
2455732.57757	&	-38.0	$\pm$	5.0	&	16.1		$\pm$	8.7	\\ 
2455732.62238	&	-25.4	$\pm$	3.8	&	5.3		$\pm$	9.3	\\ 
2455732.65737	&	-12.2	$\pm$	4.3	&	19		$\pm$	12\,\,	\\ 
2455732.69200	&	13.9	$\pm$	9.6	&	-52		$\pm$	39\,\,	\\ 
2455732.72688	&	29.2	$\pm$	3.5	&	-103.1	        $\pm$	9.2 
	\\ 
2455733.41185	&	46.1	$\pm$	7.3	&	-74.6		$\pm$	5.0	\\ 
2455733.45956	&	-2	$\pm$	22\,\,  &	-55.3		$\pm$	5.5	\\ 
2455733.50435	&	-23.6	$\pm$	9.7	&	-42.6		$\pm$	5.2	\\ 
2455733.54024	&	-31.4	$\pm$	5.3	&	-22.6		$\pm$	5.9	\\ 
2455733.57612	&	-32.5	$\pm$	9.7	&	-4.6		$\pm$	9.5	\\ 
2455733.66781	&	-41		$\pm$	17\,\,  &	14		$\pm$	14\,\,
	\\ 
2455734.03182   &	-73		$\pm$	14\,\,&	51			$\pm$   20 \,\, 
\\ 
2455734.43476	&	-108.0	$\pm$	7.2	&	79.0		$\pm$	7.4	\\ 
2455734.51579	&	-100.0	$\pm$	6.8	&	72.8		$\pm$	8.0	\\ 
2455734.55967	&	-99.2	$\pm$	9.0	&	66.5		$\pm$	8.7	\\ 
2455734.60265	&	-94.6	$\pm$	9.0	&	53.4		$\pm$	8.9	\\ 
2456086.65388	&	-108.1	$\pm$	4.2	&	82.3		$\pm$	5.1	\\ 
2456087.47573	&	82.3	$\pm$	4.5	&	-146.8	        $\pm$	5.5
	\\ 
2456087.57988	&	119.9	$\pm$	4.2	&	-193.3	        $\pm$	5.3
	\\ 
2456087.67136	&	123.6	$\pm$	4.2	&	-198.0	        $\pm$	5.1
	\\ 
2456087.73410	&	113.4	$\pm$	4.0	&	-184.5	        $\pm$	5.0
	\\ 
2456088.61205	&	-91.0	$\pm$	4.2	&	59.7		$\pm$	4.9	\\ 
2456089.57301	&	37.8	$\pm$	3.8	&	-105.8	        $\pm$	6.5
	\\ 
2456089.67823	&	88.0	$\pm$	4.1	&	-155.9	        $\pm$	4.9
	\\ 
2456089.73287	&	109.0	$\pm$	4.7	&	-181.1	        $\pm$	5.6
	\\ 
2456090.71762	&	-79.3	$\pm$	4.4	&	49.7		$\pm$	5.1	\\ 
2456091.49676	&	-71.9	$\pm$	5.4	&	39.4		$\pm$	6.3	\\ 
2456092.44932	&	37	$\pm$	15\,\,  &	-64.9		$\pm$	7.1	\\\hline 
\multicolumn {3}{c}{KPNO}\\\hline 
2456085.68447&	 79.6 $\pm$	2.9& -138.4 $\pm$	0.8\\ 
2456087.66230&	127.4 $\pm$	2.9& -198.0 $\pm$	0.7\\ 
2456087.95906&	 45.6 $\pm$	2.3& -100.9 $\pm$	1.8\\ 
2456088.70413&	-95.7 $\pm$	3.5&   75.1 $\pm$	1.9\\ 
2456091.95950&	126.3 $\pm$	5.0& -191.4 $\pm$	1.2\\ 
\hline 
\end{tabular} 
\end{center} 
\end{table}

Thirty eight spectra were obtained using the Intermediate dispersion 
Spectrograph and Imaging System (ISIS) on the William Herschel Telescope (WHT). 
The spectra were taken 2011 June 18--21 and 2012 June 7--14 with a resolving 
power of R\,$\sim$17\,000 and $\sim$22000, respectively. Calibration exposures 
using CuAr and CuNe lamps were taken prior to each 300-s exposure of 
KIC\,4544587.  Blue and red spectra were obtained using a wavelength coverage of 
$4200-4550$\,\AA\,\,and $6100-6730$\,\AA\,\,respectively. The gratings H2400B 
(blue arm) and R1200R (red arm) were used. A 0.5-arcsec slit was used to give 
Nyquist sampling on the CCD and to limit radial velocity errors due to the 
positioning of the star within the slit. The signal-to-noise obtained was 
$\sim$100 per resolution element. The data were reduced using optimal extraction 
techniques as implemented in the {\sc pamela} package \citep{Marsh1989}.  
 
\begin{figure*} 
\hfill{} 
\includegraphics[width=\hsize, height=12cm]{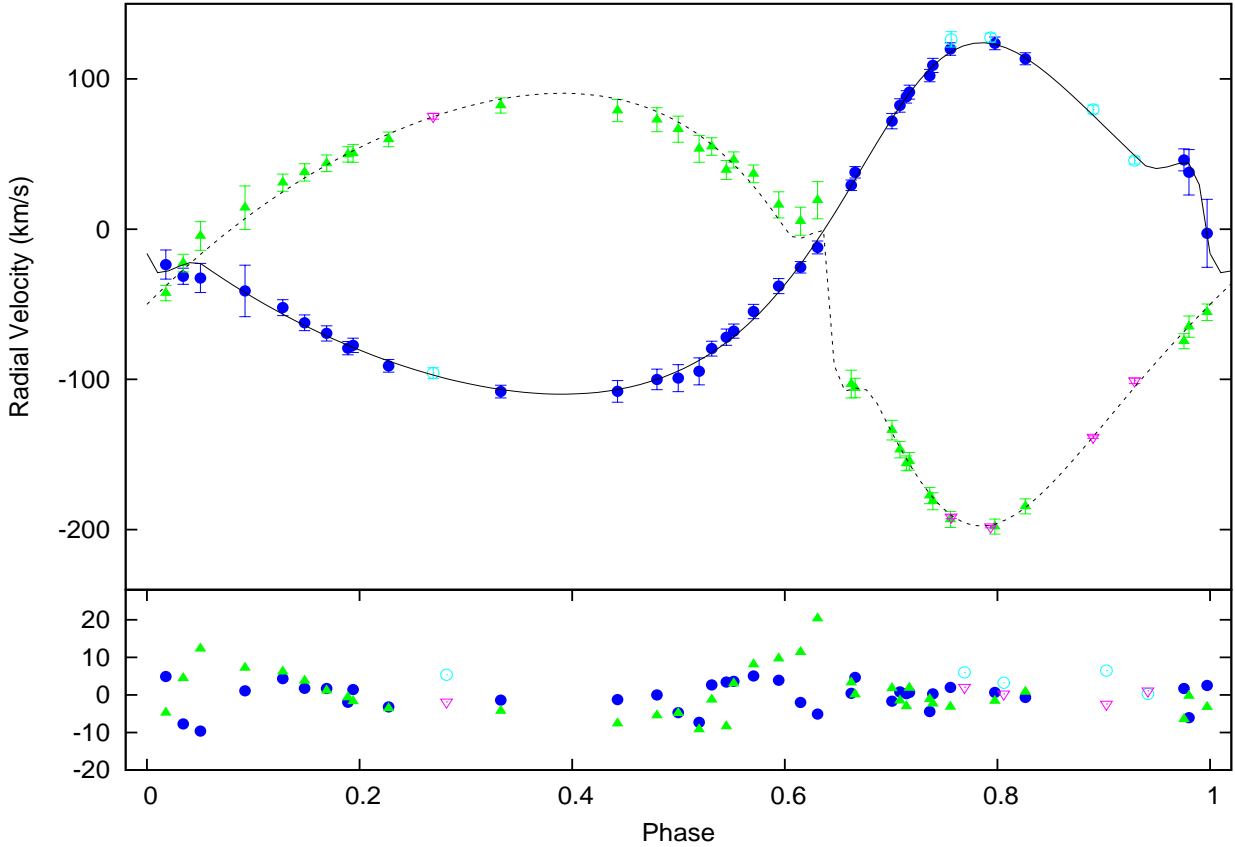}  
\small\caption{{\it Top panel}: Radial velocity curve generated from 38 spectra 
obtained using ISIS on the WHT and 5 spectra obtained using the echelle 
spectrograph on the 4-m Mayall telescope at KPNO, folded over the orbital 
period. The blue solid squares and light blue open circles represent the primary 
component from the WHT and KPNO data, respectively the green solid triangles and 
the pink open triangles represent the secondary component from the WHT and KPNO 
data, respectively, and the solid and dashed lines represent the primary and 
secondary components. The errors bars show the uncertainties in the radial 
velocity measurements. {\it Bottom panel}: The residuals from the best fit to 
the radial velocity data. } 
\label{RVcurve} 
\hfill{} 
\end{figure*} 
 
The initial radial velocity curves were determined using the 2-D cross-correlation technique as implemented in TODCOR \citep{Zucker1994} on the red and 
blue spectra together. The templates for the primary and the secondary component 
were taken from \citet{Castelli2004} model atmospheres, using T$_1$\,=\,8250\,K, 
$\log$\,$g$$_1$\,=\,4.0, [M/H]$_1$\,=\,0.0, and T$_2$\,=\,8000\,K, 
$\log$\,$g$$_2$\,=\,4.0, [M/H]$_2$\,=\,0.0, respectively. Of the 38 spectra 
taken, cross-correlation failed to produce a good radial velocity fit for 1 
spectrum. Subsequent improvement to the radial velocity curves was done by 
revising the templates according to the best fit photometric model and applying 
them to both blue and red ends: T$_1$\,=\,8600\,K, $\log$\,$g$\,$_1$\,=\,4.24, 
and T$_2$\,=\,7750\,K, $\log$\,$g$\,$_2$\,=\,4.33, respectively. A systematic 
offset slightly larger than 1$\sigma$ was found between the radial velocities of 
the red arm and the blue arm. As there is no obvious cause for this discrepancy 
each simultaneous exposure was averaged over the two arms and the discrepancy 
included in the uncertainty of the radial velocity measurements. The final 
radial velocity data have a typical 1$\sigma$ uncertainty of 
$\sim$7.3\,km\,s$^{-1}$  and are listed in Table\,\ref{tab:RVs} and depicted in 
Fig.\,\ref{RVcurve} with the best-fit radial velocity model folded on the period and zero point obtained from the light curve. 
 
\begin{figure} 
\hfill{} 
\includegraphics[width=6cm, height=\hsize,angle=270]{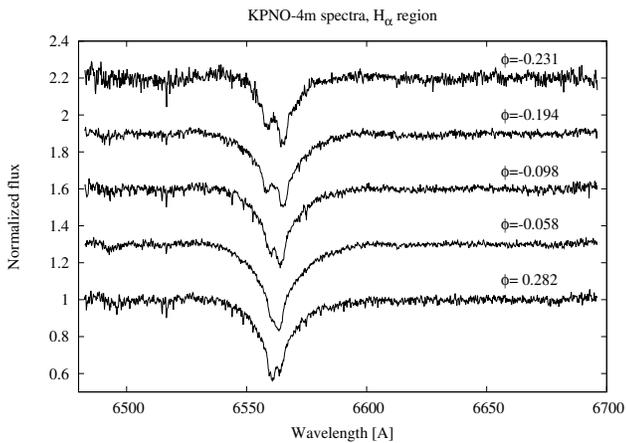}  
\small\caption{The H${}_\alpha$ region of the 5 echelle spectra acquired by 
the 4-m Mayall telescope at Kitt Peak with $R \sim 20,000$ and the 
wavelength span $4500-9000$\,\AA. The components are clearly resolved in the five spectra.} 
\label{Halpha} 
\hfill{} 
\end{figure}

Subsequently, five high-resolution spectra were taken using the echelle 
spectrograph on the 4-m Mayall telescope at KPNO with R\,$\sim$20,000 and a 
wavelength range of 4500--9000\,\AA. The data were wavelength-calibrated and 
flux-normalised as depicted in Fig.\,\ref{Halpha} where Doppler splitting is 
clearly visible. As the per-wavelength signal-to-noise ratio of the KPNO spectra is notably lower than the WHT spectra, the 2-D cross-correlation technique, TODCOR, gave significantly larger uncertainties. We consequently used the broadening function technique \citep{Rucinski1992} to determine the radial velocities for KPNO spectra. The broadening functions 
are rotational broadening kernels, where the centroid of the peak 
yields the Doppler shift and where the width of the peak is a measure of the 
rotational broadening. For the template we used a radial velocity standard 
HD\,182488, with $v_\mathrm{rot} = -21.508$\,km/s. 
 
\section{Spectral disentangling: orbit} 
\label{sec:SpecDis} 
 
We applied the technique of spectral disentangling (hereafter {\sc spd}) to 
isolate spectra for the two binary components individually \citep{Simon1994}. 
Through this technique we determined the effective temperatures of the two 
components using the Balmer lines. The medium-resolution ISIS/WHT spectra, 
described in \S\,\ref{sec:spectroscopy}, contain H$\gamma$ and H$\alpha$ lines, 
and the medium-resolution echelle KPNO spectra contain H$\beta$ and H$\alpha$ 
lines. The {\sc fdbinary}\footnote{\sf http://sail.zpf.fer.hr/fdbinary} code 
\citep{Ilijic2004}, which is based on a Fourier variant of {\sc spd} 
\citep{Hadrava1995} was first applied to the time-series of ISIS/WHT spectra 
since they are more numerous than the KPNO spectra. 
Since some of the eclipse spectra are affected by the Rossiter-McLaughlin 
effect, and the line profiles are disturbed, only out-of-eclipse spectra were 
used. This substantially reduced the number of spectra available for {\spd}, but 
the phase coverage was still adequate to suppress the undulations in the 
disentangled spectra of the components \citep{Hensberge2008}. The absence of in-eclipse spectra resulted in an ambiguity in the placement of the continuum of 
the disentangled spectra. Therefore, we performed {\sc spd} in separation mode, 
and then corrected the separated spectra for line-blocking and light dilution 
using the procedure described in \citet{Pavlovski2005}.  
 
In {\sc spd} individual component spectra are calculated simultaneously and are 
self-consistently optimised with the orbital parameters, whereby the 
determination of radial velocities is bypassed \citep{Simon1994, Hadrava1995}. 
In this sense, each individual radial velocity exposure is not optimised and as such no comparison can be made with measured radial velocities \citep{Pavlovski2010}. The orbital parameters calculated by {\sc spd} are given 
in Table\,\ref{tab:SpecDis} and represent the mean values calculated through 
disentangling 5 short spectral regions from the ISIS/WHT blue spectra, which 
cover the spectral interval from $4200 - 4600$\,\AA. Telluric lines affect the 
ISIS/WHT red spectra, which are centred on the H$\alpha$ line, thus we removed 
them manually before the application of {\sc spd}. Since only five spectra were 
available in the region of the H$\beta$ line, when using {\sc spd}, we fixed all the orbital parameters with the exception of the time of periastron.  
 
\begin{figure} \centering 
\includegraphics[width=125mm]{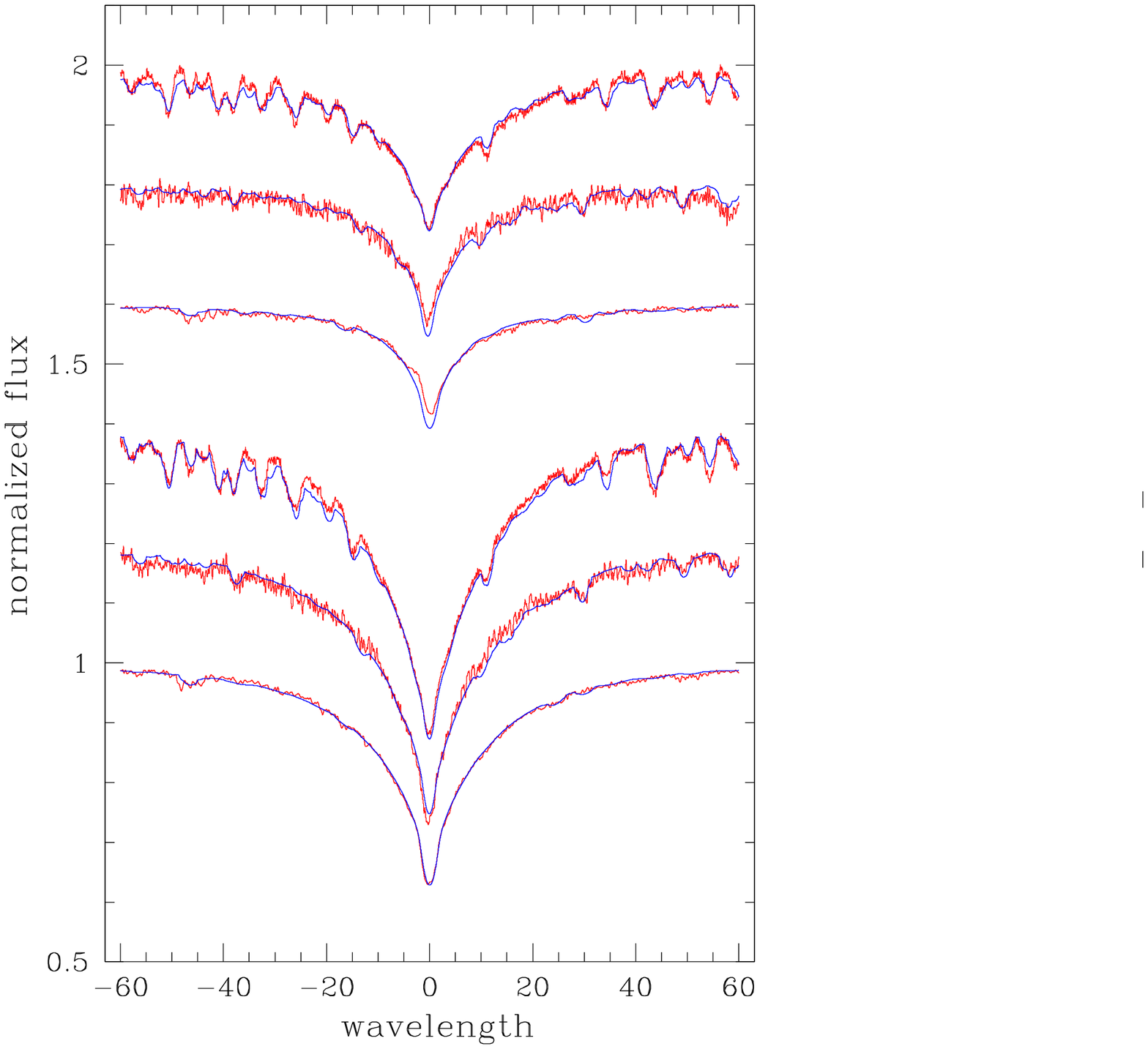} \\ 
\caption{\label{fig: fitbalmer} Comparison between disentangled 
spectra in the common continuum of the binary system (red lines, differentiable also by the noise) 
and best-fitting theoretical spectra (blue lines) for the secondary component 
(upper three spectra) and primary component (lower three spectra). The 
H$\alpha$, H$\beta$ and H$\gamma$ lines are depicted in ascending order and have 
been offset by 0.1 for clarity. } 
\end{figure}

An important outcome of {\sc spd} is an enhancement of the S/N ratio in the 
disentangled spectra, as the spectra are co-added during the {\sc spd} process. 
Due to the significant number of WHT/ISIS blue and red spectra, the S/N has 
vastly improved. However, for the KPNO spectra the gain is small due to limited 
number of spectra available for analysis. The effect of disentangling on the S/N 
ratio, for different numbers of input spectra (as well as their original S/N), 
is clearly depicted in Fig.\ \ref{fig: fitbalmer}.   
 
\begin{table} \centering 
\caption{\label{tab:SpecDis}The orbital elements of the binary system  
KIC\,4544587 derived by spectral disentangling of time series ISIS/WHT 
blue spectra.} 
\begin{tabular}{l r} 
\hline  
Parameter                                      &  SPD\\ 
\hline 
Orbital period $P$ (d)                         &  2.189094 (fix)\\  
Time of periastron passage, $T_{\rm 0}$ (BJD)       &  2455461.450(1)\\ 
Eccentricity, $e$                              &  0.288(26)\\ 
Longitude of periastron, $\omega$              & 328.5(22)\\ 
Velocity semi-amplitude $K_{\rm A}$ ($\kms$)   &  117.8(9)\\ 
Velocity semi-amplitude $K_{\rm B}$ ($\kms$)   &  145.8(10)\\ 
Mass ratio $q$                                 &  0.808(8)\\ 
\hline 
\end{tabular} 
\end{table}

\section{Atmospheric parameters} 
\label{sec:AtmoPar} 
 
Once separated, the spectra remain in the common continuum of the binary 
system, diluted by their companion's contribution to the total light of the 
system. The light ratio between the components is derived from the light curve 
solution and makes renormalisation of the individual component spectra 
straightforward.We further computed the light ratio for the Johnson $U$ (0.685), 
$B$ (0.697) and $V$ (0.667) passbands, whilst keeping all other parameters 
fixed, to determine the deviation of the light ratio as a function of 
wavelength. We note that the value derived from the light curve using the \Kep 
passband (0.670) is approximately equal to that of the Johnson $V$ band, so we 
expected that the H$\gamma$ line is most affected by our selection as its 
wavelength is furthest from the Johnson $V$ band. Consequently, with the surface 
gravities of the components known from the complementary light and RV curve 
solutions, the degeneracy between the effective temperature and the surface 
gravity can be broken. We determined the components' effective temperatures by 
fitting the re-normalised individual spectra with the synthetic theoretical 
spectra \citep{Tamajo2011}.  
 
The genetic algorithm, as implemented in {\sc pikaia} \citep{Charbonneau1990}, 
was used in the global optimisation of the code {\sc starfit} (Pavlovski et al., 
in preparation). A grid of LTE synthetic spectra was calculated using the {\sc 
uclsyn}\footnote{http://www.astro.keele.ac.uk/$\sim$bs/publs/uclsyn.pdf} code 
\citep{Smalley1994} and {\sc atlas9} model atmospheres for solar metallcity 
[M/H] = 0 \citep{Castelli1997}. The grid covers  $\Teff$ from 6\,000 to 
10\,000\,K in steps of 250\,K, and $\log$\,g from 3.50 to 4.50 in step of 
0.5\,dex.  
 
The projected rotational velocities of the components were also optimised. 
However, as convolution with the rotational kernel has little influence on the 
broad Balmer lines, we avoided simultaneous determination of the $T_{\rm eff}$ 
and $v\,\sin\,i$. Instead, we determined the $v\,\sin\,i$ of the components by 
fitting the least blended metal lines. The results are given in Table \ref{tab: 
atmospar}. As $v \sin i = 86 \pm 13$\,km\,s$^{-1}$, KIC\,4544587 has an equitorial velocity of $v_{eq}$\,$<$\,120\,km\,s$^{-1}$, below which diffusion can occur. Thus it is likely that the primary component is a metallic-lined Am star \citep{Abt2009}. 
 
\begin{table} \centering 
\caption{\label{tab: atmospar}Atmospheric parameters for the components of 
KIC~4544587, derived from a constrained optimal fit of the disentangled spectra with the light factor as a free (upper section) and fixed (middle section) parameter. For both the fitted and fixed case the projected rotational velocities (lower section) are fixed to the 
values derived by the optimal fitting of the metallic lines in the disentangled 
ISIS/WHT blue spectra.} 
\begin{tabular}{l r r} 
\hline 
Parameter  &  Primary  &  Secondary\\ 
\hline 
$T_{\rm eff}$ (K)                   &   8\,090(90)         &  7\,620(135)\\ 
$\log g$ (cgs)                      &   4.22(fixed)        &  4.23(fixed)\\ 
Light factor                        &  0.670(fixed)        &  0.330(fixed)\\ 
\hline 
$T_{\rm eff}$ (K)                   &  8\,600(100)         & 7\,750(180)\\ 
$\log g$ (cgs)                      &  4.12(2)             & 4.31(2)\\ 
Light factor                        & 0.634/0.646/0.670    & 0.366/0.354/0.330\\ 
\hline 
$v$\,$\sin$\,$i$ (km\,s$^{-1}$)     & 86.5(13)             & 75.8(15)\\  
\hline 
\end{tabular} 
\end{table}

Alongside the optimal fitting of the re-normalised disentangled spectra of 
H$\gamma$, H$\beta$, and H$\alpha$ lines separately (with the surface gravities 
and projected rotational velocities held fixed), we have also derived optimal 
atmospheric parameters in the constrained mode \citep{Tamajo2011}. In 
constrained mode the light ratio between the components is a free parameter when 
fitting for the effective temperatures. Also, the surface gravities were left to be free parameters. In the search for the optimal set of parameters, we also 
adjusted for the velocity shift between disentangled and theoretical spectra, to 
enable a slight adjustment of the continua of the disentangled spectra. 
Disentangling the Balmer lines is a difficult task due to their broadening, 
which is much larger than their Doppler shift. Moreover, when determining the 
effective temperature, the correct continuum placement is difficult because the 
Balmer lines of the primary extend over a considerable number of echelle orders, 
making the correction of the blaze and order merging somewhat uncertain. The 
optimal set of the parameters obtained when performing constrained fitting, with the light factor as both a free and fixed parameter, 
 is given in Table \ref{tab: atmospar}.

\section{Period Determination} 
\label{sec:Period}

Period analysis was performed to identify the orbital period of the binary 
system. An initial estimate was obtained by applying {\sc period04} 
\citep{Lenz2004} to the SC data from Quarter 3.2 only. {\mbox{\sc period04}} 
applies a Fourier transform to the data and uses a least-squares fit to optimise 
the amplitudes and phases. Further analysis was then performed on all the SC 
data (Quarters 3.2, 7 and 8) using {\it kephem} \citep{Prsa2011}, an interactive 
package with a graphical user interface that incorporates 3 methods: Lomb-
Scargle (LS; Lomb 1976; Scargle 1982),\nocite{Lomb1976, Scargle1982} Analysis of 
Variance (AoV; Schwarzenberg-Czerny 1989),\nocite{Schwarzenberg-Czerny1989} and 
Box-fitting Least Squares (BLS; 
Kov{\'a}cs et al. 2002),\nocite{Kovacs2002} as implemented in the {\it vartools} 
package (Hartman et al. 1998)\nocite{Hartmann1998}. Using {\it kephem}, the 
period and time of primary minimum were found interactively. The 
period was determined by dragging the mouse over a periodogram in the lower 
panel of the GUI to see how it affected the alignment of the phased data 
presented in the upper panel of the GUI. To determine an accurate period the 
zoom tool was utilised on both the periodogram and phased data. The zero point 
was then selected by dragging the primary eclipse in the top panel containing 
the phased data and align it with zero phase. The ephemeris was found to 
be:\newline 
\newline 
Min\Rmnum{1} = BJD 2455462.006137(9)+2.189094(5) $\times$ E\newline 
\newline 
\noindent 
where the values in the parentheses give the uncertainty in the previous digits. 
The uncertainties were obtained by identifying the range of values that would 
yield a visibly indistinguishable result; beyond this uncertainty range the discrepancy is notably increased. Due to apsidal motion the relative separation of the eclipses changes as a function of the rotation of the orbit. Consequently, the period specified is the anomolaus period, which is the period measured by phasing the data on one eclipse (primary eclipse), leaving the other eclipse (secondary eclipse) smeared. Although this is a small effect, the smearing could be seen over the duration of the data used in this analysis. See \S\ref{sec:orb_evol} for further discussion on the apsidal motion of KIC\,4544587.

\begin{figure*} 
\includegraphics[width=\hsize]{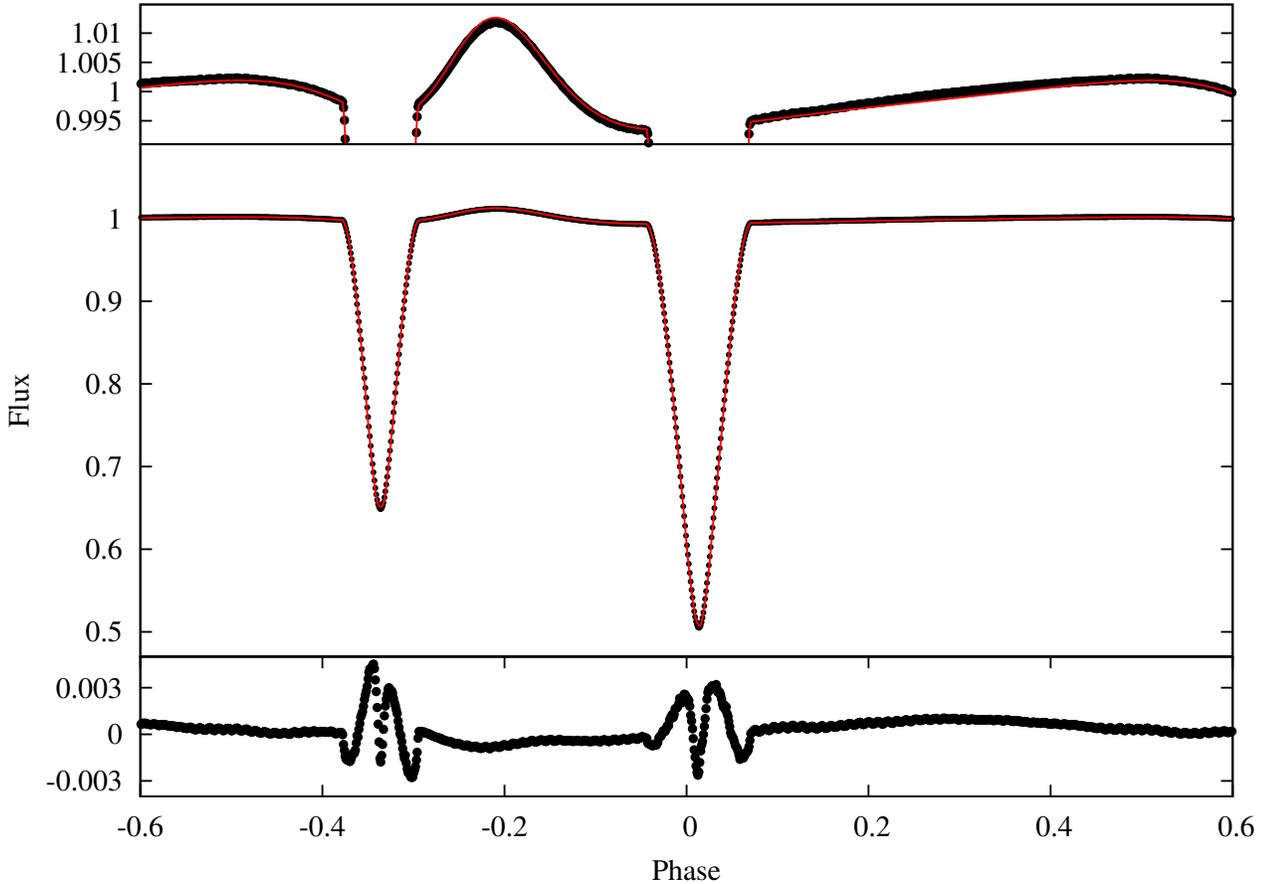}  
\small\caption{Middle panel: theoretical {\sc phoebe} model (red line) and 
observed light curve, prewhitened with the pulsation frequencies displayed in 
\S\ref{sec:Pulse} (black points) for the short cadence data of Quarters\,7 and 
8. Lower panel: the residuals (black points) of the best-fit model. Upper panel: a magnified image of the out of eclipse data and \ph model fit.} 
\label{Model} 
\end{figure*} 
 
\section{Binary Modelling} 
\label{sec:Mod} 
 
\subsection{PHOEBE} 
\label{sec:Binary} 
 
{\sc phoebe} \citep{Prsa2005} is a binary modelling package based on the Wilson-Devinney (hereafter WD) code \citep{Wilson1971,Wilson1979,Wilson2004}. {\sc 
phoebe} incorporates all the functionality of the WD code but also provides an 
intuitive graphical user interface alongside many other improvements that make 
{\sc phoebe}  highly applicable to the precise \Kep data. These include: uncertainty calculations 
through heuristical scanning algorithms (which scan parameter space by 
generating results from multiple starting points to determine the mean and 
standard deviation); the facility to phase-bin the data; updated filters for the 
various recent space missions including {\it Kepler}; the correct treatment of 
reddening; and the ability to work with a large number of data points. 
 
When modelling the data, the initial inputs were a combination of the effective 
temperatures and $\log$\,$g$ values identified through fitting the disentangled 
spectra with the light factors as a free parameter, 8600\,$\pm$\,100\,K, 7750\,$\pm$\,180\,K, 4.12 and 4.31. We elected to use the results from this mode as a single light factor does not account for the change in each component's relative light contribution for the different spectral ranges. For the initial investigation a model light curve was 
generated from the observationally constrained and estimated input parameters. 
First, the passband luminosity of the model was computed so that the out of 
eclipse flux levels were correctly positioned with respect to the observed light 
curve data. Following this the eccentricity ($e$) and argument of periastron 
($\omega$) were adjusted until the separation between the primary and secondary 
eclipses, which is proportional to $e$\,$\cos$\,$\omega$, was equal to that of 
the observed data. This also involved adjusting the phase shift to retain the 
position of the model's eclipses. Once the separation was tightly constrained, 
the phase shift, $e$ and $\omega$ were further adjusted, whilst maintaining the 
value for $e$\,$\cos$\,$\omega$, to obtain the relative widths of the primary 
and secondary eclipses, which are proportional to $e$\,$\sin$\,$\omega$. The 
combined depths and widths of the eclipses were then adjusted by altering the 
inclination and stellar potentials, respectively. 
 
\begin{table} 
\hfill{} 
\caption{ 
\label{tab:ParamFreePh} 
\small Adjusted parameters and coefficients of the best fit model to the {\it 
Kepler} light curve for Quarters\,7 and 8. The limb darkening coefficients correspond to the logarithmic limb darkening law. 
The uncertainties were determined through modelling and Monte Carlo methods, and 
concur with those obtained through fitting the disentangled spectra.The limb 
darkening coefficients were taken from the {\sc phoebe} limb darkening tables 
\citep{Prsa2011}.}  
\begin{center} 
\begin{tabular}{||l|r||} 
\hline 
Parameter   &{Values}\\ 
\hline 
Mass ratio								& 0.810(12)\\ 
Primary mass ($\Msun$), $M_1$ 			& 1.98(7)\\ 
Secondary mass ($\Msun$), $M_2$			& 1.61(6)\\ 
Primary radius (\Rsun), $R_1$			& 1.82(3)\\ 
Secondary radius (\Rsun), $R_2$			& 1.58(3)\\ 
Phase shift  					& 0.0831(3)\\ 
Semi-major axis (\Rsun), $a$  			& 10.855(46)\\ 
Orbital eccentricity, $e$				& 0.275(4)\\ 
Argument of periastron (rad), $\omega$          & 5.74(3)\\ 
Orbital inclination (degrees), $i$		        & 87.9(3)\\ 
Primary T$_{\mathrm{eff}}$ (K), $T_{1}$             & 8600(100)\\ 
Secondary T$_{\mathrm{eff}}$ (K), $T_2$	        & 7750(180)\\ 
Primary potential, $\Omega$$_1$  		& 7.09(10)\\ 
Secondary potential, $\Omega$$_2$ 		& 7.12(10)\\ 
Gamma velocity (km\,s$^{-1}$) 			& -20.13(7)\\ 
Apsidal Advance (y/cycle)			& 182(5)\\ 
Sidereal Period (d)				& 2.1890951(7)\\ 
Primary relative luminosity	  	        & 0.668(2)\\ 
Secondary relative luminosity			& 0.332(1)\\ 
Primary $\log$\,$g$ (cgs),  $\log$\,$g$$_1$     & 4.241(9)\\ 
Secondary $\log$\,$g$ (cgs), $\log$\,$g$$_2$	& 4.33(1)\\ 
Primary linear limb darkening coeff.		& 0.634\\ 
Secondary linear limb darkening coeff. 		& 0.664\\ 
Primary logarithmic limb darkening coeff. 	& 0.282\\ 
Secondary logarithmic limb darkening coeff.     & 0.268\\ 
\hline 
\end{tabular} 
\hfill{} 
\end{center} 
\end{table}

\begin{table} 
\caption{ 
\label{tab:ParamFix} 
\small Fixed parameters and coefficients for the {\sc phoebe} best-fit model to 
the {\it Kepler} light curve for Quarter\,7.  The rotation is specified as a 
ratio of stellar to orbital rotation and the fine grid raster is the number of 
surface elements per quarter of the star at the equator and coarse grid raster 
is used to determine whether the stars are eclipsing at a given phase.} 
\begin{center} 
\begin{tabular}{||l|r||} 
\hline 
Parameter & Values\\ 
\hline 
Third light                          		& 0.0\\ 
Orbital Period (d)               	    	& 2.189094(5)\\ 
Time of primary minimum (BJD)		        & 2455462.006137(9)\\ 
Primary rotation                     		& 1.83\\  
Secondary rotation                   		& 1.83\\  
Primary Bolometric albedo        		& 1.0\\  
Secondary Bolometric albedo   	            	& 1.0\\ 
Primary gravity brightening             	& 1.0\\ 
Secondary gravity brightening           	& 1.0\\ 
Primary fine grid raster             		& 90\\  
Secondary fine grid raster                	& 90\\  
Primary coarse grid raster              	& 60\\  
Secondary coarse grid raster            	& 60\\  
\hline 
\end{tabular} 
\end{center} 
\end{table}

Once an initial model had been generated, the differential corrections algorithm 
was applied in an iterative process to obtain an accurate fit to the light curve 
data. Once the model was tightly constrained, the radial velocity curves were 
incorporated to fit the mass ratio, gamma velocity and the projected semi-major 
axis. As the photometric light curve contains essentially no information about 
these parameters for a detached system, the fit was performed on the radial 
velocity curves independently. This avoids improper weighting due to the vastly different number of data points between the different types of curves. Once the best 
fit solution had been achieved for these parameters the differential corrections 
algorithm was applied to the light curve for all other parameters specified in 
Table\,\ref{tab:ParamFreePh}. 
 
When generating the model we assumed pseudo-synchronous stellar rotation after 
\citet{Hut1981}, which was determined to be 1.87 times the orbital period. 
pseudo-synchronous rotation is indicative of the rotational velocity of the 
stellar components at periastron. We also fixed the orbital period since {\it 
kephem} is more appropriate for period determination than the differential 
corrections algorithm. Due to the low contamination and following the analysis 
of the pixel level data we assumed no third light in the system.  
 
\begin{figure*} 
\hfill{} 
\includegraphics[height=\textwidth,angle=270]{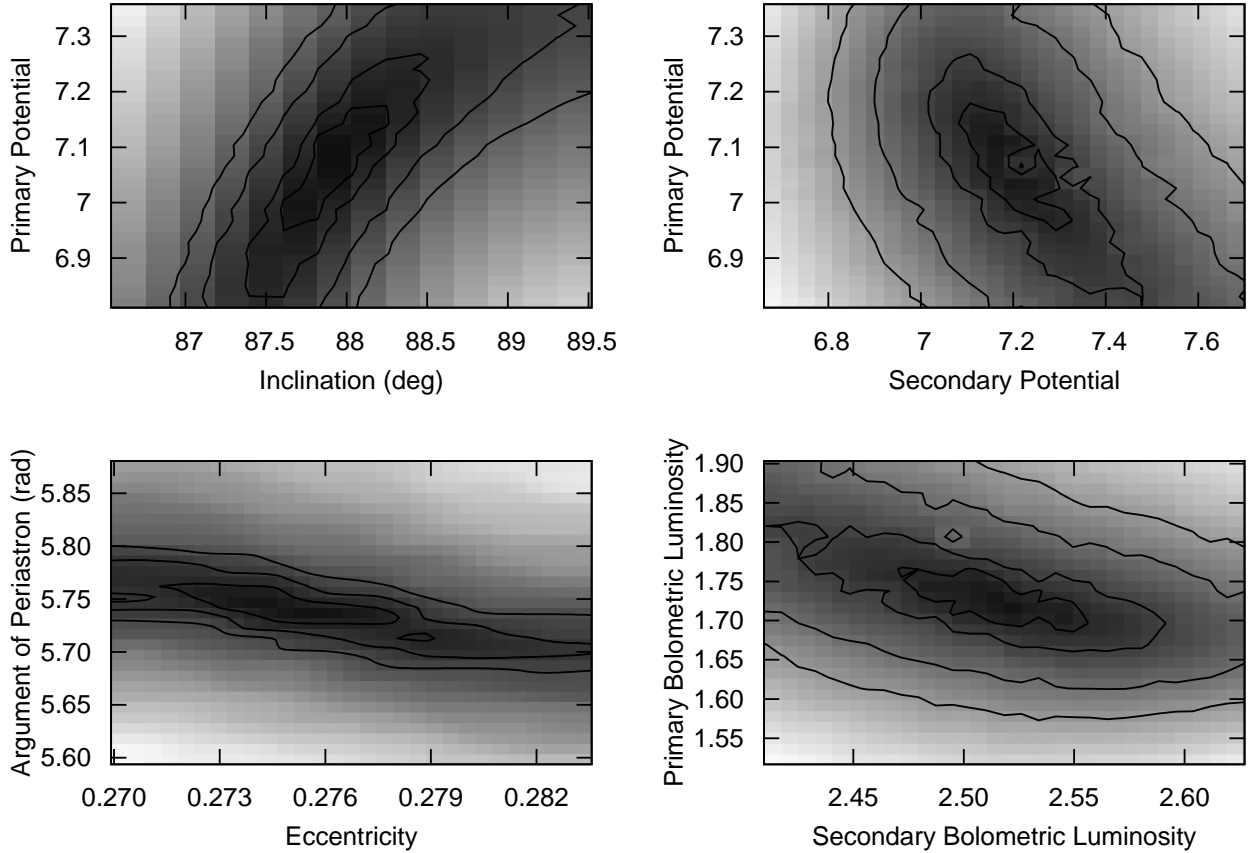}  
\small\caption{Density maps showing the distribution of results from the Monte 
Carlo simulations for the most correlated parameters: potential of the primary 
vs. inclination (top left), primary vs. secondary potential (top right), 
argument of periastron vs. eccentricity (bottom left) and the luminosity of the 
primary vs. the luminosity of the secondary (bottom right). The contours 
represent the uncertainty in terms of standard deviation, with the innermost 
contour representing the 1$\sigma$ uncertainty and subsequent contours 
representing increments of 1$\sigma$.} 
\label{fig:MC} 
\hfill{} 
\end{figure*}

\begin{figure} 
\hfill{} 
\includegraphics[width=\hsize, height=\hsize,angle=270]{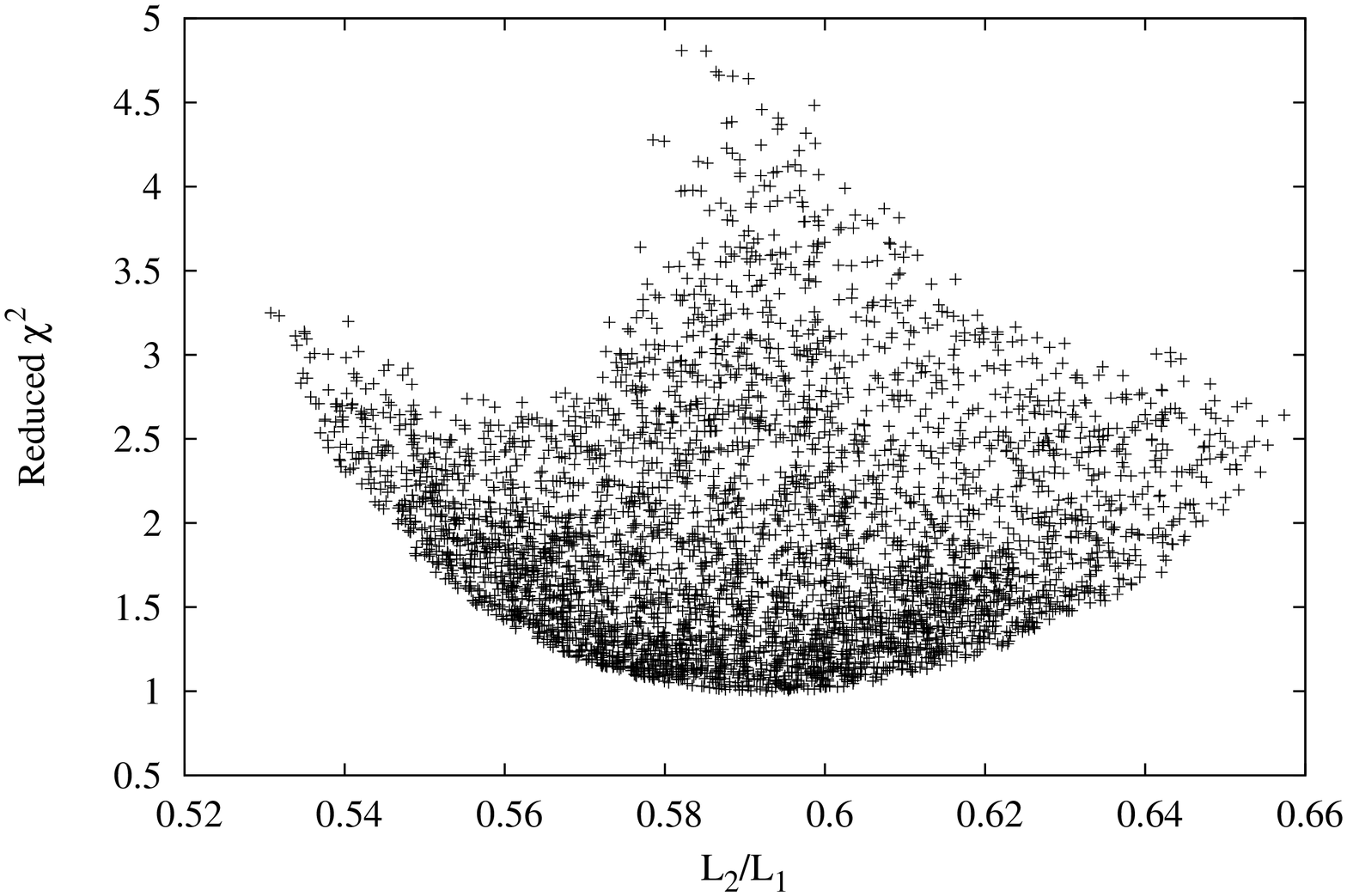}  
\small\caption{A scatter plot showing the spread of results obtained from the 
Monte Carlo simulation for the luminosity ratio as a function of $\chi$$^2$. A clear boundary with a definite 
minimum can be seen, which signifies that the luminosity ratio is unambiguous.} 
\label{fig:Scatter} 
\hfill{} 
\end{figure}

When modelling a binary system with one or more pulsating components (where the 
pulsations occur on the time scale of the orbit), multiple iterations are 
required so that the data are thoroughly prewhitened, leaving only the binary 
signature. This enables the orbital characteristics to be modelled correctly 
without interference from the stellar pulsations. The method used involved 
subtracting the computed orbital model from the original observed data, 
subsequent frequency analysis on the residual data, and finally, the removal of 
the identified pulsations from the original data. This method is only viable 
when the pulsations can be considered as perturbations, which is the case for 
KIC\,4544587. What remains is a light curve predominantly free of pulsations for 
subsequent binary modelling. Three iterations were required when modelling 
KIC\,4544587, with subsequent iterations having negligible effect. The fitted 
and fixed parameters, and their corresponding values for our best fit model, can 
be found in Tables\,\ref{tab:ParamFreePh} and \ref{tab:ParamFix} respectively. 
 
The model obtained for KIC\,4544587, as seen in Fig.\,\ref{Model}, still shows 
some systematic discrepancies in the residuals during primary and secondary 
eclipse. These 
discrepancies arise from a combination of 1) the existence of pulsations that 
are commensurate with the orbital period and 2) the precise nature of the \Kep 
data. As some of the pulsations are commensurate with the orbital period, they 
occur at precisely the same time each orbit. During eclipse phase, however, the 
relative flux from the pulsating component either increases or decreases, 
dependent on which star is being eclipsed. This introduces a change in the 
amplitude of the pulsation during eclipse phase that manifests itself in the 
residuals of the model. Additionally, the highly precise \Kep data have 
highlighted the inadequate treatment of parameters such as limb darkening, 
stellar albedo and the incomplete treatment of surface discretization  
\citep{Prsa2011}, which have previously been considered satisfactory. Currently 
efforts are being made towards improving the models to account for the physics 
that has previously been omitted (Pr{\v s}a et al. 2013, in prep.). However, until this 
major task, which is outside the scope of this paper, is completed, these 
systematics are unavoidable when generating a binary model of the \Kep data and 
thus are accounted for in the uncertainties attributed to the fitted parameters.   
 
Uncertainty estimates were obtained using a combination of formal errors, 
generated by fitting all the parameters simultaneously using {\sc phoebe}, and 
those determined through Monte Carlo heuristic scanning. A scan of the parameter 
space was undertaken for the most correlated parameters using Monte Carlo 
methods; the results of the Monte Carlo simulations can be found in 
Figs\,\ref{fig:MC} and \ref{fig:Scatter}. The Monte Carlo simulations perturbed 
the solutions of the best fit model by a predefined amount (5\%) in order to 
identify the spread of possible results and their corresponding $\chi$$^2$ 
values. The $\chi$$^2$ values for each solution were then mapped out into 
confidence intervals which serve as the uncertainty estimates. From the density 
maps the optimum values of the correlated parameters displayed on the axses can 
be seen by identifying the combination with the lowest $\chi$$^2$ value. The 
1$\sigma$ uncertainty values are determined by considering the spread of 
the innermost contour. 
 
\begin{figure} 
\hfill{} 
\includegraphics[width=\hsize, height=7cm]{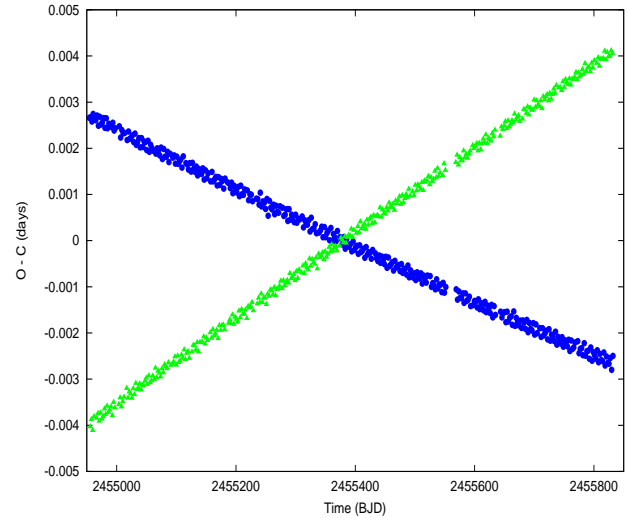}  
\small\caption{Depicted are the eclipse timing variations, which demonstrate the deviation of the observed 
primary (blue circles) and secondary (green triangles) eclipses from the model 
as a function of time.  The average period was selected for the standard model 
and thus it can be seen that both eclipses deviate from each other. The opposing 
direction of the primary and secondary eclipses is suggestive of classical 
apsidal motion. With a longer time base it is expected that the trend would 
appear more sinusoidal. The standard model was generated using 4 polynomials of 
order 10, which was found to be the optimal order to reduce the scatter.} 
\label{fig:ETV} 
\hfill{} 
\end{figure}

\subsection{Orbital Evolution} 
\label{sec:orb_evol}
 
Apsidal motion is the rotation of the elliptical orbit about the centre of mass 
\citep{Claret1993}, which can be caused by the presence of a tertiary component 
or through the gravitational interactions occurring between the binary 
components. Using {\sc phoebe} we determined the rate of apsidal advance for 
KIC\,4544587 to be 0.04306\,(2)\,rad\,y$^{-1}$. This equates to a full rotation 
of the orbit in 182(5)\,y. Fig.\,\ref{fig:ETV} shows the eclipse timing variations of KIC\,4544587, and depicts the 
primary and secondary eclipses moving linearly in opposite directions, which is 
suggestive of classical apsidal advance caused by tidal disortions. 
\citet{Gies2012} modelled the eclipse timing variations of 41 eclipsing 
binaries, including KIC\,4544587, and concluded that these variations are a 
combination of apsidal motion and tidal interactions.

The parameters determined by {\sc phoebe} are indicative of an eccentric short-period system with two components in close proximity ($\sim$4.4 \Rsun\ surface-to-surface, at periastron). Although the periastron distance is small, we still do not expect 
any mass transfer as the potentials of the two components far exceed the 
potential at L1 ($\sim$4.8), as specified by {\sc phoebe}.  
 
Eccentric short period binary systems are uncommon due to the rapid rate of 
circularisation (10$^{7}$\,y) compared with the timescale of stellar evolution. 
\citet{Zahn1975} theorised that orbital circularisation occurs due to the 
radiative damping of tidally excited oscillations in the stellar outer envelope. 
For this reason the eccentric short period nature of KIC\,4544587 is either the 
result of tidal capture, recent formation of the system, the presence of a 
tertiary component or a consequence of its resonant pulsations. As the \Kep 
field does not contain any prominent star forming regions, it is not expected 
that KIC\,4544587 is a newly formed binary system. Furthermore, as the system is 
formed from two intermediate mass main sequence stars, it is also unlikely that 
the system has undergone tidal capture. Thus the eccentric nature of 
KIC\,4544587 is likely a consequence of either a third body or the system's 
extreme tidal interactions (see \S\ref{sec:Tidal} for a discussion of tidal 
resonance).

\section{Pulsation Characteristics} 
\label{sec:Pulse} 
 
\begin{table} 
\hfill{} 
\begin{center} 
\caption{ 
\label{tab:Freqs} 
The identified pulsation frequencies and their corresponding amplitudes and 
phases. The values in parentheses give the 1$\sigma$ uncertainty in the previous 
digit. The uncertainty in the amplitude is 0.004\,$\times$\,10$^{-3}$ relative 
flux units.} 
\begin{tabular}{l r r r} 
\hline 
 Designation & Frequency & Amplitude & Phase\\ 
 &(d$^{-1}$)& (flux\,$\times$10$^{-3}$) & (rad)\\ 
\hline 
{\it f}$_{1}$ = {$\nu$}$_{orb}$ 	&0.45681(1)	&	1.001	&0.746(3)\\ 
{\it f}$_{2}$ = 4{$\nu$}$_{orb}$	&1.82710(1)	&	0.593	&0.3822(4)	\\ 
{\it f}$_{3}$ 	&2.01124(1)	&	0.561	&	0.0758(5)	\\ 
{\it f}$_{4}$ = 3{$\nu$}$_{orb}$ 	&	1.37041(1)	&	0.520 &	0.3514(5)
	\\ 
{\it f}$_{5}$	&3.46822(1)	&	0.373&	0.9864(7)	\\ 
{\it f}$_{6}$ 	&48.02231(4)	&	0.329	&	0.547(2)	\\ 
{\it f}$_{7}$ = 7{$\nu$}$_{orb}$    	&	3.19760(2)	&	0.244	&
	0.301(1)	\\ 
{\it f}$_{8}$ 	&	41.37020(5)	&	0.236	&	0.892(3)	\\ 
{\it f}$_{9}$ 	&	44.84695(6)	&	0.181	&	0.780(3)	\\ 
{\it f}$_{10}$ 	&	0.12546(3)	&	0.164	&	0.996(2)	\\ 
{\it f}$_{11}$   	&	46.19662(8)	&	0.152	&	0.990(4)	\\ 
{\it f}$_{12}$	&	0.12721(4)	&	0.140	&	0.396(2)	\\ 
{\it f}$_{13}$ = 97{$\nu$}$_{orb}$	&	44.30982(9)	&	0.134	&	0.776(4)
	\\ 
{\it f}$_{14}$ = 2{$\nu$}$_{orb}$    	&	0.91388(4)	&	0.133	&
	0.026(2)	\\ 
{\it f}$_{15}$	&	48.04449(19)	&	0.122	&	0.331(9)	\\ 
{\it f}$_{16}$ = 10{$\nu$}$_{orb}$       &	4.56792(4)	&	0.116	&
	0.532(2)	\\ 
{\it f}$_{17}$ = 8{$\nu$}$_{orb}$	&	3.6545(5)	&	0.106	&	0.055(2)
	\\ 
{\it f}$_{18}$	&	39.54280(11)	&	0.106	&	0.984(6)	\\ 
{\it f}$_{19}$	&	1.61186(5)	&	0.103	&	0.897(3)	\\ 
{\it f}$_{20}$ 	&	43.44756(12)	&	0.101	&	0.733(6)	\\ 
{\it f}$_{21}$	&	44.81796(12)	&	0.099	&	0.088(6)	\\ 
{\it f}$_{22}$ = 9{$\nu$}$_{orb}$	&	4.11122(6)	&	0.093	&	0.813(3)
	\\ 
{\it f}$_{23}$	&	46.58340(13)	&	0.092	&	0.056(7)	\\ 
{\it f}$_{24}$	&	0.04089(6)	&	0.091	&	0.829(3)	\\ 
{\it f}$_{25}$ 	&	1.58541(7)	&	0.078	&	0.545(3)	\\ 
{\it f}$_{26}$	&	38.22668(16)	&	0.076	&	0.109(8)	\\ 
{\it f}$_{27}$	&	44.29902(21)	&	0.054	&	0.80(1)	\\ 
{\it f}$_{28}$	&	44.36118(29)	&	0.052	&	0.70(1)	\\ 
{\it f}$_{29}$	&	40.05372(23)	&	0.051	&	0.30(1)	\\ 
{\it f}$_{30}$	&	46.67401(23)	&	0.051	&	0.10(1)	\\ 
{\it f}$_{31}$	&	44.75638(24)	&	0.049	&	0.68(1)	\\ 
{\it f}$_{32}$	&	47.95373(26)	&	0.045	&	0.69(1)	\\

\bottomrule 
\end{tabular} 
\hfill{} 
\end{center} 
\end{table} 
 
\begin{figure*} 
\includegraphics[width=\hsize]{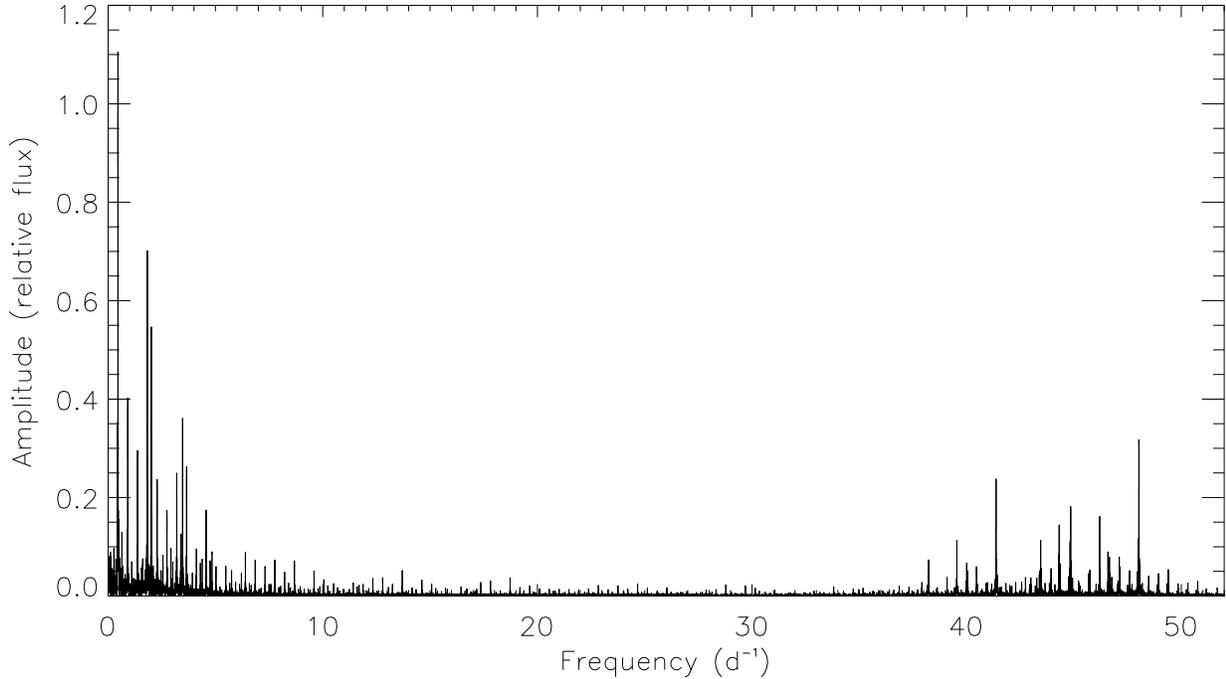}  
\small\caption{An amplitude spectrum of the residual, eclipse-masked data for 
the short cadence data of Quarters 7 and 8. Here the high frequency p-mode 
regime is clearly separated from the lower frequency g-mode regime.} 
\label{FT} 
\end{figure*} 
 
The light curve of KIC\,4544587 demonstrates clear pulsations in two regions of 
the frequency spectrum: with periods on the order of days and periods on the 
order of 30\,min, both of which can be seen in the light curve. We used {\sc 
period04} and our own codes to generate a frequency spectrum of the residual 
data (the detrended data with the orbital fit subtracted), which can be seen in 
Fig.\,\ref{FT}. We also performed eclipse-masking by removing the data points 
occurring during eclipse phases to remove any residual binary information from 
the light curve. In the Fourier transform, the gaps created in the data manifest 
themselves in the window pattern, with peaks separated from the real peak by the 
orbital frequency. Although this is not ideal, masking is highly important for 
the identification of resonantly excited modes, which is a crucial aspect of the 
analysis of KIC\,4544587. Without removing the aforementioned points, the 
systematics would have presented themselves as frequencies at multiples of the 
orbital frequency, identical to the signature of tidally excited modes, thus 
masking was required to differentiate between these two possibilities.  
 
{\sc period04} incorporates a least squares fitting technique to simultaneously 
generate amplitudes and phases for all the identified frequencies. For an 
assumed background level of 40\,$\mu$mag we report the frequencies with 
amplitudes of 3$\sigma$ or more. The prominent frequency peaks were identified 
in two regions, $0-5$\,d$^{-1}$ and $30 - 50$\,d$^{-1}$, which correspond to 
g\,modes and p\,modes, respectively, although the lowest frequency peak 
(f$_{24}$\,=\,0.04089(6)\,d$^{-1}$) is possibly due to remaining instrumental 
effects.  
 
The high frequency, high overtone p-mode frequencies are typical for a \DS 
star of temperature similar to that of the primary star, which is towards the 
hotter, blue edge of the instability strip. An estimate of the radial overtones 
of the modes can be made from the pulsation constant, $Q$, defined by the 
period-density relation: 
 
\begin{equation} 
P \sqrt \frac{\rho}{\rho_\odot} = Q, 
\end{equation} 
 
\noindent {which can be rewritten in the form} 
 
\begin{equation} 
\log Q = -6.454 + \log P + 0.5 \log g + 0.1 M_{\rm bol} +\log T_{\rm eff}. 
\end{equation} 
 
From the latter relation and the fundamental parameters given in Table~6 we find 
for the frequency range $30 - 50$\,d$^{-1}$ for the primary star $0.017 > Q > 
0.012$, and for the secondary star $0.021 > Q > 0.015$. These are indicative of 
radial overtones in the range $3 \le n \le 5$ \citep{Stellingwerf1979}. 
Pulsation in higher radial overtone p\,modes such as these (as opposed to 
pulsation in the fundamental and first overtone modes) is more typical of hotter 
$\delta$\,Sct stars, hence suggests that the p\,modes arise in the primary star.  
 
We applied a Fourier transform up to the SC Nyquist frequency but did not find 
any peaks beyond 48.04449(19)\,d$^{-1}$. Once the frequencies specified in 
Table\,\ref{tab:Freqs} had been prewhitened, an amplitude excess still remained 
in both the p-mode and g-mode regions. However, as we could not be certain that 
further detections were real, we did not continue to extract modes beyond this 
point. 
 
We identified 31 frequencies, 14 in the lower frequency g\,mode region and 17 
p\,mode region, in the frequency spectrum between 0.1\,d$^{-1}$ and 49\,d$^{-1}$ 
(see Table\,\ref{tab:Freqs}). Of the 14 g-mode frequencies we identified 8 that 
are multiples of the orbital frequency (see\,\S\ref{sec:Tidal}).  
 
The remaining g\,modes are either \GD pulsations, most likely from the secondary 
component, which is in the \GD instability strip, or non-resonant tidally driven 
modes, as predicted by \citet{Weinberg2012}. Currently we are unable to 
differentiate as both outcomes have identical signatures, although as the 
secondary is in the \GD instability strip we would expect it to pulsate with 
intrinsically excited g\,modes.  
 
\subsection{Tidal Interactions and Combination Frequencies} 
\label{sec:Tidal} 
 
\begin{figure} 
\hfill{} 
\includegraphics[width=\hsize]{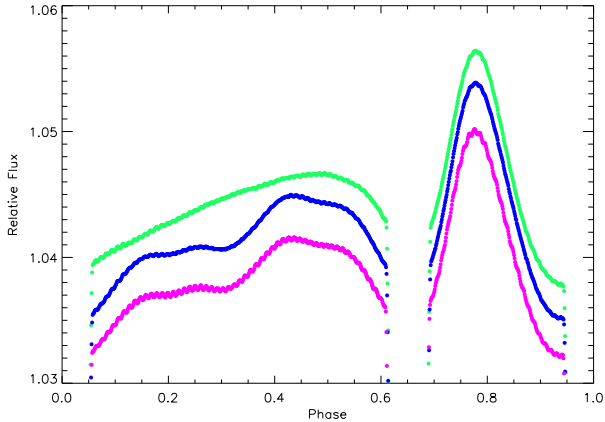}  
\small\caption{A magnified image of the phase binned \Kep photometric SC light 
curve of Quarters 7 and 8 with no frequencies prewhitened (pink, bottom curve), all the 
frequencies {\it except} those that are harmonics prewhitened (blue, middle curve) and {\it 
all} the frequencies prewhitened (green, top curve). The blue and pink light curves only 
demonstrate a minimal difference as all non-commensurate pulsations are 
cancelled out when the data is phase binned. The only explanation for the 
remaining variation in the blue and pink light curves is that they are 
resonantly excited pulsations. The light curves have been offset by 
0.03\,relative flux units for clarity.}  
\label{fig:comparison} 
\hfill{} 
\end{figure}

Tidally excited modes are stellar pulsations that have been excited by the 
tidal forces of the companion star. A prime example of this is KOI-54 
\citep{Welsh2011}. In a binary system with an eccentric orbit, when a stellar 
eigenfrequency is close to a multiple of the orbital frequency, a near resonance 
occurs that causes an increase in oscillation amplitude (relative to non-resonant modes). 
The signature of a tidally excited mode is an oscillation frequency at a multiple of the 
orbital frequency. We identified 8 frequencies in the g-mode region that are multiples 
of the orbital frequency. We expect that these are tidally excited l= 2 modes, however, as 
the eclipses are only partial, we are unable to use the in-eclipse data to determine mode 
angular degrees or which modes belong to which star. However, as the eclipses are only 
partial, we are unable to use the in-eclipse data to determine precisely which modes belong 
to which star.  
 
Previously believed to be short-lived, \citet{Witte1999} demonstrated that the 
duration of tidal resonance can be prolonged through resonant locking. Resonant 
locking is a result of the change in stellar spin, due to the exchange of 
angular momentum in the system as the orbit evolves, causing variations in the 
eigenfrequencies of the stellar components. This, combined with the change in 
orbital period of the system causes a coupling between the newly resonant 
eigenfrequency and the orbital frequency, which increases the probability of 
observing this intriguing phenomenon significantly. 
 
During the identification of the frequencies the data were masked so that the 
Fourier transform was only applied to the out-of-eclipse data. Consequently, it 
is unlikely that the presence of these frequencies in the Fourier transform can 
be completely attributed to an inadequate orbital solution. More convincingly, 
Fig.\,\ref{fig:comparison} shows a magnified image of the out-of-eclipse phase 
binned data of Quarters 7 and 8 with no frequencies removed (pink, bottom curve), all the 
identified frequencies {\it except} the orbital harmonics removed (blue, middle curve) and 
{\it all} the identified frequencies removed (green, top curve). The pink light curve is 
visibly thicker than the blue light curve because the non-commensurate 
pulsations are still present, although have essentially cancelled out. {\it We 
have no explanation for these variations other than that they are resonantly 
excited modes}. We have also ruled out spots as the cause of the light curve 
variations as spots are commensurate with the rotational frequency of the star 
and not the orbital frequency -- our rotational velocity measurements are 
consistent with pseudo-synchronous velocity, thus we would expect peaks in the 
Fourier transform at multiples of 1.87 times the orbital frequency if the 
variations were caused by spots. 
 
\begin{figure*} 
\hfill{} 
\includegraphics[width=\hsize, height=11cm]{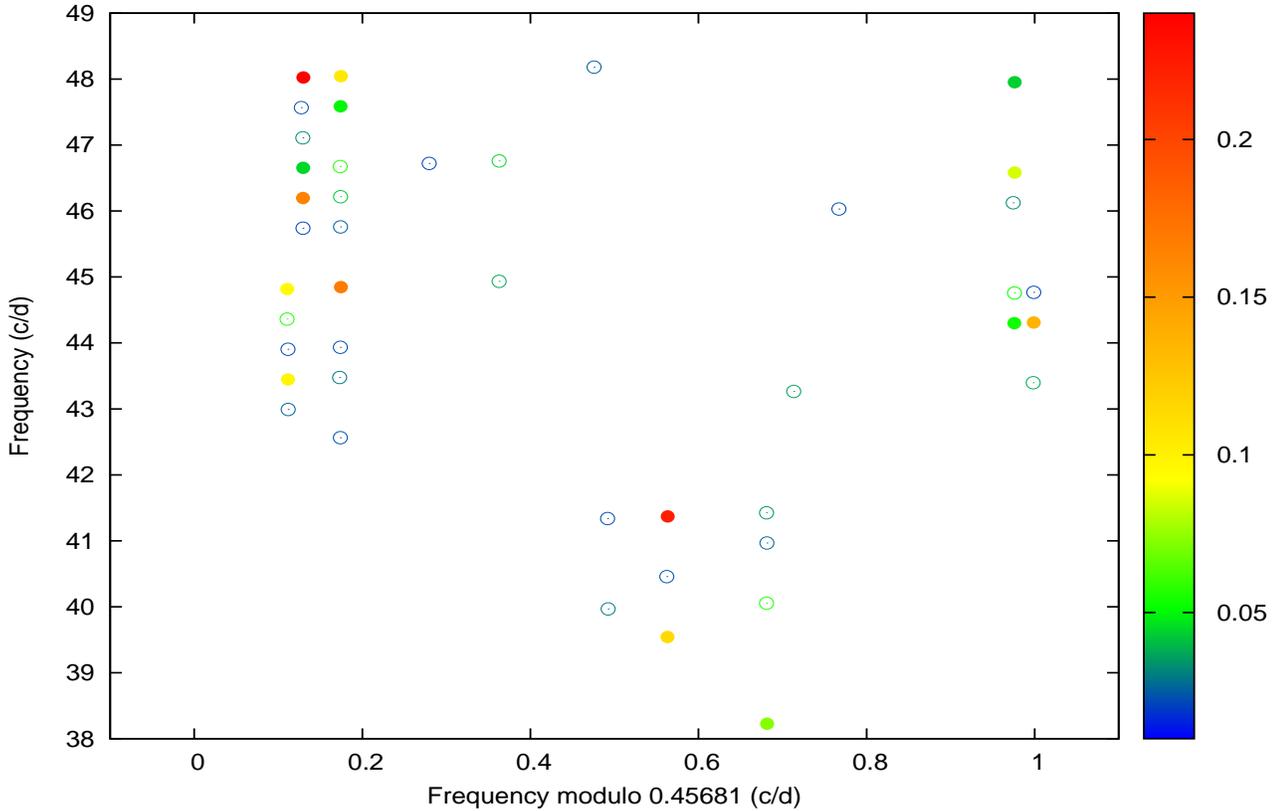}  
\small\caption{An echelle diagram of the p-mode frequencies modulo the orbital 
frequency using the SC \Kep data of Quarters 7 and 8. The points are coloured in 
terms of their amplitude in units of relative flux $\times$\,10$^{-3}$ (see the 
key at the right of the figure). The filled circles are p-mode values taken from 
Table\,\ref{tab:Freqs} and the open circles are represent frequencies with 
amplitudes in the region $0.02 - 0.04 \times 10^{-3}$ relative flux units, which 
are below our predefined confidence limit of 3$\sigma$. The uncertainty in the 
frequencies is smaller than the points, thus not depicted. A high pass filter 
was applied to the g-mode region prior to the identification of the p\,modes to 
remove any possible window pattern.} 
\label{fig:pmode} 
\hfill{} 
\end{figure*}

The p\,modes were analysed using the unmasked residual data from Quarters 7 
and 8. We generated an echelle diagram \citep{Grec1983}, modulo the orbital frequency, of the p-
mode frequencies to identify any regular spacings (cf. 
Fig.\,\ref{fig:pmode}). We prewhitened all g\,modes prior to the 
identification of the p\,modes to avoid any crosstalk from window pattern. In 
Fig.\,\ref{fig:pmode} the filled circles represent the frequencies in 
Table\,\ref{tab:Freqs} and the open circles represent frequencies with 
amplitudes in the region $0.02 - 0.04 \times 10^{-3}$ relative flux units. The 
latter are not reported in our table as they are below our predefined confidence 
limit, although here they highlight the vertical groupings which indicate that 
many of the p\,mode frequencies are split by the orbital frequency. Our working 
hypothesis is that the highest amplitude peak in a vertical group is the self-excited p\,mode and the remaining p\,modes in that group are the product of non-linear coupling between the self-excited p\,modes and tidally induced g\,modes. 
To our knowledge this effect has not previously been observed and is considered 
an important tool for identifying g\,modes in the Sun \citep{Chapellier2012}. In 
our case, this deduction suggests that one of the stars is pulsating in both 
p\,modes and g\,modes, information that we could not have determined otherwise. 
An alternative hypothesis, suggested by \citet{Weinberg2013}, states that 
through non-resonant three wave interactions the dynamical tide can excite 
daughter p-mode and g-mode waves, however, we refrain from discussing the physical nature 
of these modes at this time\footnote{After this manuscript was submitted, we realized that the g modes whose frequencies are not multiples of the orbital harmonics (i.e. $f_3$, $f_5$, $f_{19}$, and $f_{25}$ from Table 8) are indicative of non-linear tidal processes. Specifically, rather than having frequencies at orbital harmonics, these g-modes have frequencies that {\it sum} to orbital harmonics. Note that $f_3 + f_5 = 12 \nu_{orb}$, and $f_{19}+f_{25}=7 \nu_{orb}$. These combination frequencies suggest that these modes are excited by parametric three-mode resonance, as detailed in Weinberg et al. 2012 \citep{Papaloizou1981}%(see also Papaloizou \& Pringle 1981, Wu \& Goldreich 2001). 

However, the non-linear driving mechanisms may be different for the two pairs of modes listed above. In the language of Weinberg et al. 2012, the excitation of $f_{19}$ and $f_{25}$ may be due to non-linear driving by the dynamical tide. Essentially, this is the standard three-mode coupling in which a parent mode non-linearly drives pair of daughter modes whose frequencies sum to that of the parent mode, as observed in the KOI-54 system (Fuller \& Lai 2012, Burkart et al. 2012). In this case, the parent mode is the dynamical tide at $f_7 = 7 \nu_{orb}$, which is dominated by a nearly resonant g mode.  The origin of $f_3$ and $f_5$ likely can not be explained by this mechanism because there is no visible parent mode at $f_3 + f_5 = 12 \nu_{orb}$. Instead, these modes are likely excited via non-linear driving by the equilibrium tide. In this case, the "parent" mode is the component of the equilibrium tide that oscillates at 12 times the orbital frequency (which is is dominated by the f mode rather than a g mode). These findings further substantiate our pulsational models which show that we do not expect to see \GD modes in either componet. }.

Combination modes created from the same type of mode, however, have been 
observed in numerous stars including the \DS stars FG\,Virginis 
\citep{Breger2005} and KIC\,11754974 (Murphy, 2012 submitted), and the white 
dwarf star GD\,358 \citep{Winget1994}. \citet{Brickhill1983} determined that the 
likely cause of combination frequencies is nonlinear interactions related to 
convective turn-over timescales. Changes in the convective zone during the 
pulsation cycle cause a change in the amount of flux attenuation. This distorts 
the stellar shape and causes the pulsations to deviate from pure sinusoids 
generating combination frequencies in the Fourier transform. 
 
\citet{Wu2001} demonstrated that non-linear interactions could help with mode 
identification because non-linear mode coupling will only occur between specific 
modes and different modes generate different amplitude combinations, \eg 
$l$\,=\,2 modes generate larger amplitude combination frequencies than $l$\,=\,1 
modes. One of the most challenging requirements of modelling $\delta$ Sct stars 
is mode identification, thus combination frequencies could be key identifiers to 
obtaining a true asteroseismic model of the multitude of modes presented by this 
fascinating object. 
 
\subsection{Pulsation frequency modulation: the FM effect}  
\label{sec:FM} 
 
Recently \cite{sk2012} have shown that pulsating stars in binary orbits have 
frequency multiplets split by the orbital frequency in their amplitude spectra. 
This is a simple consequence of the light travel time effect causing the 
pulsation phases to be periodically modulated by the orbital motion as seen by 
the observer. When the effect is large enough to be measured, radial velocities 
can be measured directly from the light curve without the need for spectroscopic 
radial velocities. \cite{sk2012} demonstrate this by deriving the mass function 
from photometric data alone for the K-K binary and A star in the complex 
multiple system KIC\,4150611 to better precision than has been possible with 
spectroscopic radial velocities.  
 
In the case of KIC\,4544587 this effect is not measurable, even at the extremely 
high precision of the {\it Kepler} data. Consequently we are not able to 
complement our spectrally defined radial velocity points with a photometrically 
defined radial velocity curve. However, we are able to conclude that the FM 
(frequency modulation) signature is not present in our frequency spectrum and 
thus does not interfere with our frequency analysis. \cite{sk2012} characterise 
the phase modulation with a parameter, $\alpha$, given by  
 
\begin{eqnarray} 
	\alpha 
	= 
	{{\left(2\pi G{\rm M}_\odot\right)^{1/3} }\over{c}} 
	\left({{m_1}\over{{\rm M}_\odot}}\right)^{1/3} q(1+q)^{-2/3}  
	{{P_{\rm orb}^{2/3}}\over{P_{\rm osc}}} \sin i 
\end{eqnarray} 
 
\noindent 
where $P_{\rm osc}$ is the pulsation period, $P_{\rm orb}$ is the orbital 
period, $m_1$ is the mass of the primary star, and $q = m_1/m_2$ is the mass 
ratio. For KIC\,4544587 the exact value of $\alpha$ depends on which star is 
considered to be pulsating, which is not yet known, but $\alpha \sim 0.04$, in 
either case, since $q \sim 1$. For a value of $\alpha$ this low, \cite{sk2012} 
show that, to first order, a frequency triplet split by the orbital frequency is 
expected for each pulsation frequency, where the amplitude ratio of the side 
peaks to the central peak is given by 
\begin{eqnarray} 
	{{A_{+1} + A_{-1}}\over{A_n}} \approx \alpha , 
\end{eqnarray} 
where $A_{+1}$, $A_{-1}$ and $A_{n}$ represent the amplitudes of the peaks at  
$f_n + \nu_{\rm orb}$, $f_n - \nu_{\rm orb}$, and $f_n$, respectively, where 
$\nu_{\rm orb} = 1/P_{\rm orb}$ is the orbital frequency. In addition, for this 
low value of $\alpha$ the side peaks have essentially the same amplitude.  
 
Table\,\ref{tab:Freqs} shows that the highest amplitude p\,mode in KIC\,4544587, 
$f_{6} =  48.0240$\,d$^{-1}$, has an amplitude of 329\,$\mu$mag. Therefore, we 
expect the  orbital sidelobes generated by the light travel time effect to have 
amplitudes for this  best case of about 7\,$\mu$mag, which is below the limit of 
detection in our  data, but may ultimately be detectable with a more extensive 
{\it Kepler} data set.

\subsection{Stellar Evolution and Pulsation Models} 
 
To estimate the frequency content expected for stars similar to 
KIC\,4544587, we calculated stellar evolution and pulsation models for single 
spherically symmetric non-rotating stars using the mass, radius, and effective 
temperature constraints obtained from binary modelling and spectroscopic 
analysis. The models use the stellar evolution and pulsation codes described in 
\citet{Guzik2000}, including an updated version of the Iben (1963, 1965) 
\nocite{Iben1963, Iben1965} stellar evolution code with OPAL 
\citep{Iglesias1996} opacities, \citet{Alexander1994} low-temperature opacities, 
the Grevesse-Noels (1993)\nocite{Grevesse1993} solar mixture, and the Pesnell 
(1990)\nocite{Pesnell1990} linear nonadiabatic stellar pulsation codes.  We did 
not include diffusive helium, element settling or convective overshooting.  
 
We do not know the interior abundances or ages of these stars. However, we 
attempted to find evolved models of the same initial composition and age that 
matched the constraints of the two stars in KIC\,4544587. The models that were 
the closest to satisfying these criteria have helium abundance $Y = 0.27$, 
metallicity $Z = 0.017$ and mixing length/pressure scale height 1.90, very near 
the values calibrated for the Sun for these evolution and pulsation codes, and 
the Grevesee-Noels solar abundance mixture (see, e.g., Guzik $\&$ Mussack, 
2010\nocite{Guzik2010}).  The age of the models is approximately 235\,My.  The 
observational constraints, plus the model parameters for two sets of models are 
given in Table\,\ref{tab:Seismology}. 
 
\begin{figure} 
\hfill{} 
\includegraphics[width=\hsize]{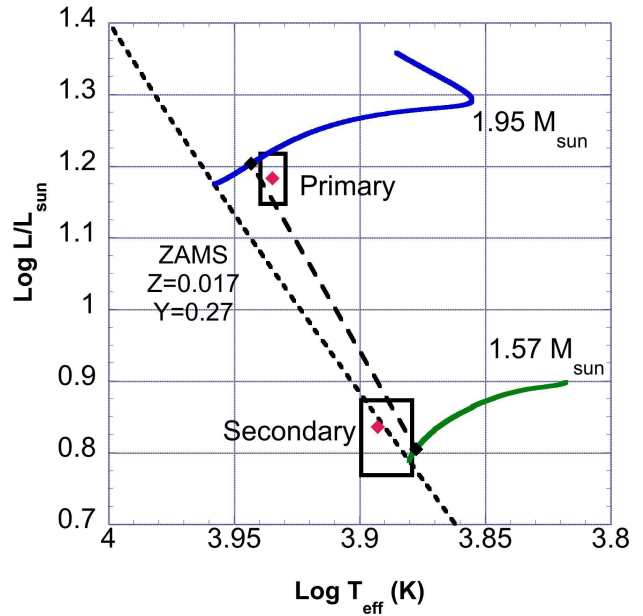}  
\small\caption{A Hertzprung-Russell Diagram for the models of 
Table\,\ref{tab:Seismology}. The boxes outline the parameter space for the 
observationally derived primary and secondary components. The short-dashed line 
is Zero-Age Main Sequence position for stellar models with $Z=0.017$, $Y=0.27$.  
Also shown are evolutionary tracks for a 1.95-M$_{\odot}$ (blue) and 1.57-
M$_{\odot}$ (green) model. The two models with the same age and composition 
(column 3 of Table\,\ref{tab:Seismology}) are closest to the observational 
constraints are connected by the long-dashed line.  The red diamonds mark the 
best fit models for each star (column 4 of Table\,\ref{tab:Seismology}) that do 
not have exactly the same age and composition.}  
\label{fig:HRD} 
\hfill{} 
\end{figure}

Fig.\,\ref{fig:HRD} shows the evolution tracks for stars of these masses and the observational constraints. The two coeval models miss the observational 
constraints by a small amount: it is difficult to find a secondary with a small-
enough radius, and a primary that is cool enough, for the same age. To gauge the magnitude of the problem, we also searched for single-star best fit models.  We found that we could fit the primary parameters 
exactly using a model with a higher metallicity (0.023), with age 171\,My; the 
higher metallicity allowed a slightly higher mass and produced the desired lower effective temperature.  For the secondary, we needed to reduce the metallicity 
to 0.016 to reduce the radius; the desired radius of 1.43 $\Rsun$ is reached at 
the zero-age main sequence.   
 
\begin{table} 
\hfill{} 
\begin{center} 
\caption{ 
\label{tab:Seismology} 
Stellar properties derived from observations, and stellar evolution models. $Y$ 
and $Z$ are initial helium and element mass fractions, respectively.  The first 
set of models (column 3) has the same age and composition, while column 4 shows 
parameters for the best-fit models that do not have the same age and 
abundances.} 
\begin{tabular}{l r r r } 
\hline 
Parameters & Observations & Coeval & Single Best-Fit \\ 
& & Models & Models\\ 
\hline 
\multicolumn{4}{l}{Primary}\\\hline 
Mass (M$_{\odot}$)		&1.98(7)	&1.95		&1.98\\ 
Radius ($\Rsun$)		&1.76(3)	&1.73		&1.76\\ 
T$_{\mathrm{eff}}$ (K)		&8600(100)	&8774		&8604\\ 
Y                               &               &0.27           &0.28\\ 
Z 				&  	        &0.017          &0.023\\ 
Age (My)			&    	        &235		&171\\ 
\hline 
\multicolumn{4}{l}{Secondary}\\ 
\hline 
Mass (M$_{\odot}$)		&1.60(6)		&1.57		&1.60\\ 
Radius ($\Rsun$)		&1.43(2)		&1.48		&1.43\\ 
T$_{\mathrm{eff}}$ (K)	        &7750(180)	        &7543		&7813\\ 
Y                               &                       &0.27           &0.27\\ 
Z 				&  	                &0.017  	&0.016\\ 
Age (My) 		        & 	                &235		& 0 (ZAMS)\\ 
\bottomrule 
\end{tabular} 
\hfill{} 
\end{center} 
\end{table}

We calculated the $l$\,=\,0, 1, and 2 nonadiabatic pulsation frequencies for the 
models in Table\,\ref{tab:Seismology}.  For the models with age 235\,My, we find 
that the 1.95-$\Msun$ model is predicted to have only one unstable $\delta$ Sct-
type frequency for $l$\,=\,0 (7$^{th}$ overtone, 59.4\,d$^{-1}$),  $l$\,=\,1, 
(56.3\,d$^{-1}$) and $l$\,=\,2 (58.9\,d$^{-1}$). The 1.57-$\Msun$ model for the 
secondary is predicted to have unstable $\delta$ Sct-type frequencies in a very 
wide range, 21 to 90\,d$^{-1}$ for the $l$\,=\,0 (radial) modes (Fundamental 
through 10$^{th}$ overtones),  21 to 93\,d$^{-1}$ for the $l$\,=\,1, and  22 to 
96\,d$^{-1}$ for the $l$\,=\,2 modes.  Overall, the p-mode frequency content of 
KIC\,4544587 seems to correspond better to that of the secondary, contrary to the findings of the Q-value equation, but these 
calculations do not rule out tidal excitation of modes in the primary.  
 
The predicted unstable frequency content for the best-fit non-coeval models in 
Column 4 of Table\,\ref{tab:Seismology} is similar, with the primary showing 
just a few p\,modes for each degree $l$\,=\,0, \,1 and \,2 between 46 and 
61\,d$^{-1}$, a little higher but overlapping the p-mode frequency range 
observed in KIC\,4544587.  For the 1.60-M$_{\odot}$ secondary, the range of 
unstable p\,modes is a little smaller than for the 1.57-M$_{\odot}$ higher-metallicity model in column 3 of Table\,\ref{tab:Seismology}, and shows eight unstable modes between 22 and 75\,d$^{-1}$ for each degree $l$\,=\,0, \,1 and 
\,2.   
 
We also examined the possibility of unstable g-mode frequencies for these models 
in the observed range of $1.5 - 3.5$\,d$^{-1}$. While we can find many g\,modes 
for either $l$\,=\,1 or $l$\,=\,2 in this frequency range, all of the modes are 
stable.  The models have envelope convection zones that are quite shallow, with 
a temperature at the base of $50,000-60,000$\,K, near or a little hotter than 
the region of second helium ionization. At this location, the convective 
timescale at the convection zone base is shorter than the pulsation period, so 
\GD g\,modes are not expected to be pulsationally unstable via the convective 
blocking mechanism. Verification of this conclusion using a pulsation code with 
a time-dependent convection treatment would be worthwhile (see, e.g., Grigahcene 
et al. 2004, Dupret et al. 2005). On the other hand, in either star, g\,modes 
consistent with those observed could be tidally or stochastically excited. 
 
While they are useful to guide expectations for intrinsic pulsation frequencies, 
these single star non-rotating pulsation and evolution models are  not adequate 
for an asteroseismic analysis. Tidal effects and rotation will distort the 
stars, and so the pulsation modes observed will not correspond to those 
calculated assuming spherical symmetry.  Mode coupling will alter the observed 
frequencies. The surfaces of these stars are separated by only 4.3\,$\Rsun$ at 
periastron.  It is unlikely, but possible, that mass transfer occurred during an 
earlier evolution stage. In addition, tidal forces can cause mixing of hydrogen 
into the stellar core that would slow the evolution compared to that of single 
stars and alter the internal structure \citep{Liakos2012}.

\section{Summary and Conclusions} 
\label{sec:Conc} 
 
We have presented the \Kep photometric and ground based spectroscopic model of 
KIC\,4544587, a detached eclipsing binary system with p-mode  and g-mode 
pulsations, apsidal motion, tidally excited modes and combination frequencies. 
The SC data of Quarters\,7 and 8 were used in the binary modelling and pulsation 
analysis of KIC\,4544587 with the exception of modelling the apsidal motion and 
eclipse timing variations where all available Quarters were used. Radial 
velocity curves have been incorporated into the binary model, which were 
generated from 38 spectra obtained using ISIS on the WHT and five spectra using 
the echelle spectrograph on the 4\,m Mayall telescope at KPNO. 
 
The binary model was created using {\sc phoebe} in an iterative process where 
the pulsations were identified in the residuals of the orbital fit and 
subsequently prewhitened to leave only the binary signature for modelling 
purposes. We were able to obtain a reasonable, but not completely ideal binary 
model fit. Primarily this is due to the resonant pulsations in KIC\,4544587, 
which have periods commensurate with the orbital period and do not diminish when 
phasing the data. However, it is also because of the inadequate treatment of 
certain physical parameters (most prominently gravity brightening, limb 
darkening and albedo) in the binary modelling process, highlighted by the 
precise nature of the \Kep data. Addressing this deficiency is a work in progress (Pr{\v s}a et al. 2013, in prep., Degroote et al. 2014, in prep.).  
 
A best fit model was obtained and uncertainty estimates were determined using a 
combination of formal errors and Monte Carlo simulations, which were used to 
determine uncertainties for parameters that are highly correlated. The 
distributions obtained in these simulations  demonstrated minimal degeneracy, 
attesting to the uniqueness of the  obtained binary solution. From the binary 
model fit we determined the fundamental parameters of the stellar components. 
These include the mass and radius of the primary \DS component, $1.98 \pm 
0.07$\,$\Msun$ and $1.82 \pm 0.03$\,\Rsun, respectively, and the mass and radius 
of the secondary component, $1.61 \pm 0.06$\,$\Msun$ and $1.58 \pm 0.03$\,\Rsun. 
We also determined that the system has rapid apsidal motion, $182 \pm 5$\,y per 
cycle, which may be partially attributable to the resonant pulsations. 
 
The binary characteristics were subsequently separated from the inherent 
pulsations and 31 periodicities were identified, 14 in the g-mode region and 17 
in the p-mode region. Of the 14 g-mode pulsations, 8 were found to have 
frequencies that are multiples of the orbital frequency, therefore we conclude 
that the majority of these are tidally excited pulsations. Seventeen p-mode 
frequencies were identified in the residuals, many of which demonstrate 
separations that are multiples of the orbital frequency. Our current hypothesis 
is that these are combination modes, formed through the non-linear interactions 
between p\,modes and g\,modes. The stellar pulsation models predict many more 
unstable p\,modes for the secondary component than the primary, so it is 
possible that these p\,modes originate in the secondary  component; however, the 
secondary could also have a few unstable p\,modes.  The pulsation models show 
that neither star has a convection zone deep enough to produce unstable 
g\,modes, at least via the convective blocking mechanism.    The g\,modes, 
however, could also be tidally driven and originate with either the primary or 
secondary.  Further investigation into the nonlinear mode interactions and tidal 
excitation of the pulsation modes will require modelling of pulsations in 
tidally distorted rotating stars, and is beyond the scope of this paper. 
 
\section{Acknowledgements} 
The authors express their sincere thanks to NASA and the \Kep team for allowing 
them to work with and analyse the \Kep data, making this work possible. The \Kep 
mission is funded by NASA's Science Mission Directorate. This work was also 
supported by the STFC (Science and Technology Funding Council). We would also 
like to thank the RAS for providing grants which enabled KH's attendance at 
conferences and thus enabled the development of collaborations and the 
successful completion of this work. AP acknowledges support through NASA Kepler 
PSP grant NNX12AD20G. The research leading to these results has received funding 
from the European Research Council under the European Community's Seventh 
Framework Programme 
(FP7/2007--2013)/ERC grant agreement n$^\circ$227224 (PROSPERITY), as 
well as from the Research Council of KU Leuven grant agreement GOA/2008/04. We 
acknowledge the observations taken using the 4-m Mayall telescope at the NOAO, 
survey number \#11A-0022, and the Isaac Newton Group of Telescopes for the use 
of the William Herschel Telescope (WHT). The WHT is operated on the island of La 
Palma by the Isaac Newton Group in the Spanish Observatorio del Roque de los 
Muchachos of the Instituto de Astrofisica de Canarias. The authors would also 
like to thank Susan Thompson and William Welsh for their comments 
and suggestions.

%     R E F E R E N C E S 
 
\bibliographystyle{mn2e} 
\bibliography{StarRef_2}  
\label{lastpage}

\end{document}